\documentclass[twocolumn]{aastex631}
\usepackage{bm}
\usepackage{amsmath}
\def\la{\hbox{\raise.35ex\rlap{$<$}\lower.6ex\hbox{$\sim$}\ }}
\def\ga{\hbox{\raise.35ex\rlap{$>$}\lower.6ex\hbox{$\sim$}\ }}

\def  \postrevisionbf{}

\def\St{{\rm {St}}}

\def\beq{\begin{equation}}
\def\eeq{\end{equation}}
\def\beqa{\begin{eqnarray}}
\def\eeqa{\end{eqnarray}}

\def\order#1{{\cal O}\left({#1}\right)}

\definecolor{darkgreen}{HTML}{25911E}

\definecolor{purple}{rgb}{0.7,0.0,0.7}

\received{\today}
\submitjournal{Astrophysical Journal}
\shorttitle{Streaming Instability in Global Turbulence}
\shortauthors{Estrada \& Umurhan}
\graphicspath{{./}{figures/}}

\begin{document}

\title{Formation of the First Planetesimals via the Streaming Instability in Globally Turbulent Protoplanetary Disks?}

\author[0000-0002-1132-5594]{Paul R. Estrada} \affiliation{Space Sciences Division, Planetary Systems Branch, NASA Ames Research Center,  Mail Stop 245-3, Moffett Field, CA 94035, USA}
\correspondingauthor{Paul R. Estrada}
\email{paul.r.estrada@nasa.gov}
\author[0000-0001-5372-4254]{Orkan M. Umurhan}
\altaffiliation{New Horizons Science Team, Co-Investigator}
\affiliation{Space Sciences Division, Planetary Systems Branch, NASA Ames Research Center,  Mail Stop 245-3, Moffett Field, CA 94035, USA}
\affiliation{SETI Institute, 389 Bernardo Way, Mountain View, CA 94043, U.S.A.}
\affiliation{Cornell Center for Astrophysics and Planetary Sciences, Cornell University, Ithaca, NY 14853, USA}
\affiliation{Department of Earth \& Planetary Sciences, University of California Berkeley, Berkeley, CA 94720, USA}
\altaffiliation{Lecturer 2022-2024 academic years.}


\defcitealias{Umurhan_etal_2020}{U+20}
\defcitealias{Yang_etal_2017}{Y+17}
\defcitealias{Carrera_etal_2015}{C+15}
\defcitealias{Li_Youdin_2021}{LY21}
\defcitealias{Estrada22_paper2}{E+22}
\defcitealias{Schafer_Johansen_2022}{SJ22}
 
\begin{abstract}

Using self-consistent models  
of turbulent particle growth in an evolving protoplanetary nebula of solar composition we find that recently proposed local metallicity and Stokes number criteria necessary for the streaming instability to generate gravitationally bound particle overdensities
are generally not approached anywhere in the disk during the first million years, an epoch in which meteoritic and observational evidence strongly suggests that the formation of the first planetesimals and perhaps giant planet core accretion is already occurring.
\end{abstract}


\section{Introduction} \label{sec:intro}

Understanding how the first planetesimals were
made constitutes perhaps the single most critical linchpin in deciphering the history of planet formation in the solar nebula and beyond.  In addition to providing the fundamental building blocks of planetary nuclei envisaged in the core accretion model, clarifying the means by which planetesimals formed should go toward also explaining the extensive chemical and lithological mixing that took place in the early solar disk as evidenced by the meteorite record \citep[e.g.,][]{Zolensky2006,Krot2009,Joswiak2012,Kita2013,Kruijer2017,Zanda2018,Marrocchi2019,Simon2019}. 
\par
In this regard, a commonly held view is that planet formation proceeded differently before and after the formation of Jupiter's core \citep{Nanne2019,Jacquet2019}. The general perception is that the emergence of the first planetary cores hastens the 
formation of gravitationally bound planetary structures throughout the disk: that is, a sufficiently massive perturbing body in the form of a giant planet core (putatively in the $10-20$ M$_\oplus$ range) helps to drive tidal torques within the disk gas that lead to the formation of gaps \citep[e.g.][]{LP79,Dangelo_etal_2003,Paardekooper2018}, which in turn aid in the trapping of particles at gap edges \citep[e.g.,][]{Paardekooper2006,Rice2006,Desch_etal_2018,Draz2019}.  Moreover, the planetary core, embedded in the ambient disk gas, can benefit from an enhanced solids capture rate 
through the 
process of pebble accretion \citep[e.g.,][]{Ormel2010,Lambrechts2012,Chambers2014}, in which particle-gas drag helps to focus particles onto a growing core.  Yet, assembling these cores relies on prior formation of planetesimals. 
\par
Interpretations of the meteorite record suggest that the first planetesimals -- {\it i.e.}, the array of $40-100$ km scale gravitationally bound mass bodies that presumably went into building planetary cores -- were formed as early as $\sim 0.5$ Ma (millions of years) and as late as $\sim 3-4$ Ma after CAIs \citep{Kita2012,Connelly2012,Schrader2017,Kruijer2017}. Furthermore, the combination of the above-mentioned chemical and lithological mixing as well as observations of line-broadening in protoplanetary disks \citep[e.g.,][``pp-disk'' hereafter]{Teague2016,Flaherty2017,Flaherty2018,Dullemond2018}  
suggest that the early solar nebula was turbulent; indeed, it is increasingly thought that the early stages of evolution of pp-disks are at least weakly-to-moderately turbulent in the regions where particle growth is of the greatest interest \citep[$\lesssim 100$ au, see e.g., reviews by][]{Turner2014,Lyra_Umurhan_2019,Lesur_etal_2022}. 

However, 
particle growth by sticking (typically mm-cm sized for compact pebbles) in 
pp-disk gas flows encounters 
growth ``barriers''  
and loss to the central star via radial drift before planetesimals can ever form \citep[see][and references therein]{Estrada_etal_2016}. 
Fractal, porous aggregates can survive longer 
in the pp-disk, but under the same nebula conditions eventually they too suffer the same fate as their compact counterparts  \citep[][hereafter E+22]{Estrada22_paper2}.
Thus, it has been argued that some mechanism must come into play that collects these growth-frustrated 
particles into gravitationally bound multi-km sized structures -- objects that are ``born big'' \citep{Morbidelli2009}.
\par
Three mechanisms that could help ``leap-frog" across these growth barriers have been cited in the recently released Planetary Science Decadal Survey \citep{Planetary_Science_Decadal_Survey_2022},
which include particle trapping by large scale coherent gaseous vortices
\cite[e.g.][and references therein]{Raettig_etal_2021}, and turbulent concentration \cite[e.g.][and references therein]{Hartlep_etal_2020}.  The third and current leading candidate mechanism is the Streaming Instability (SI, hereafter), in which the relative velocity between a pressure-free particle component and a weakly pressure-supported rotating gaseous fluid brings about a gas-drag mediated momentum exchange resonance that leads to high densities in the particle field \citep{Youdin_Goodman_2005,Squire_Hopkins_2018a,Squire_Hopkins_2018b,Lesur_etal_2022}. High resolution shearing box simulations under certain conditions predict the SI leads to the efficient formation of a resolved distribution of self-gravitating particle overdensities 
\citep[e.g.,][]{Simon_etal_2017}
with resulting rotational profile statistics that appear consistent with the angular momentum distribution observed in the cold-classical Kuiper Belt Object (KBO) population \citep{Nesvorny_etal_2019}.
\par
Several studies have examined the question of under what local conditions does the SI operate efficiently to produce gravitationally bound overdensities \citep[][hereafter C+15, Y+19, LY21, respectively]{Carrera_etal_2015,Yang_etal_2017,Li_Youdin_2021}. These studies addressed this matter using the results of direct numerical simulations of shearing box calculations of globally laminar disks covering a range of values in 
the disk metallicity $Z$ and particle Stokes number $\St$. In these simulations, any turbulence $\alpha$ is {\it self-generated} in the midplane-settled particle layer with its magnitude dependent on the choice of $Z$ and $\St$.

It is generally reported that the occurrence of strong overdensities by the SI needs the ratio of the midplane solids-to-gas mass density $\epsilon$ to hover around or greatly exceed unity.  
For a range in Stokes numbers ($0.001 < {\rm St} < 1$) this condition appears to require that at the very minimum $Z$ exceed 0.015.  Recently \citetalias{Li_Youdin_2021} report from their globally laminar models that the minimum $Z$ needed for occurrence drops to as low as 0.005 for $\St > 0.1$.  However, \citetalias{Li_Youdin_2021} also suggest that the minimum $Z$ ought to increase with and when including external sources of $\alpha$ as well after accounting for model turbulence calculations reported in \citet{Gole_etal_2020}.
\par
Whether the efficient operation of the SI can be attained in realistic models of the early solar nebula has not yet been established.  There are conflicting indications as to whether or not the SI is effective or even viable in disks experiencing global turbulence \citep[][the last two of these hereafter U+20 and SJ22, respectively]{Gole_etal_2020,Chen_Lin_2020,Schafer_etal_2020,Umurhan_etal_2020,Schafer_Johansen_2022}
with turbulent intensities $10^{-5}<\alpha <10^{-3}$ \citep{Lyra_Umurhan_2019,Lesur_etal_2022}, strongly depending upon the thermal cooling time \citep[e.g.,][]{Richard_etal_2016,Manger_etal_2021}. Indeed, even in the laminar case, the settling dust layer generated 
$\alpha$ is large enough to thwart the SI for St $\order{0.01}$: the recent high resolution simulation of \citet{Carrera_Simon_2022} to model the action of the SI in a globally laminar model 
shows that while midplane settled mm grain-sized dust layers become turbulent at the level of $\alpha \sim 6\times 10^{-5}$, they do not lead to appreciable overdensities even though the local $Z$ ($\approx 0.03$) and St ($\approx 0.015$) conditions of the simulations are expected to be strongly SI active based on the above cited occurrence studies \citepalias{Carrera_etal_2015,Yang_etal_2017,Li_Youdin_2021}.  Preliminary models of the 
results of global disk evolution within the first Ma with turbulent particle growth suggest that the conditions for the efficient operation of the SI are not met according to the analytical theories of the SI subject to turbulence \citep[\citetalias{Umurhan_etal_2020};][]{Chen_Lin_2020} mainly because when the St of the particles or aggregates grow to large enough values, it comes at the expense of a corresponding decrease in the local values of the metallicity due to rapid radial drift \citep[\citetalias{Estrada22_paper2};][]{Estrada22_paper3}.
\par
The purpose of this study is to ask the question: {\it Are the conditions under which the SI can produce gravitationally bound particle overdensities actually met in the first million years of evolution of globally turbulent protoplanetary nebulae, an epoch in which evidence strongly suggests the first planetesimals formed?} 
In Section \ref{sec:SIoccur} we briefly review the state of the science regarding SI, and discuss the so called SI occurrence diagrams. In Section \ref{sec:globmod} we introduce the global nebula evolution models we use in this work. In Section \ref{sec:results}, we discuss the results of our analyses and in Section \ref{sec:summary} we summarize our main conclusions.

\section{SI occurrence, a review}\label{sec:SIoccur}
 
The SI is initiated when a relative stream occurs between the 
particle phase and the gas.  There are several instances of such relative streaming configurations, but the most common setting is when there is a global radial gas pressure gradient which induces the gas to orbit the star slower than the particles.  This situation leads to momentum exchange between the two phases causing particles to steadily spiral inward, and the gas outward 
\citep[e.g., the steady solutions of][]{Nakagawa_etal_1986}.
The strength of this relative flow is governed by the Stokes number defined in the Epstein regime (where the particle size $a$ is smaller than the gas molecular mean free path) as
\beq
\St = \frac{3 m_{\rm{p}}}{4 \rho c A}\Omega = \frac{\rho_{\rm p}}{\rho} \frac{a}{c}\Omega,
\eeq
with 
the second equality corresponding to compact spherical particles where the {\postrevisionbf{particle}} mass-to-cross-sectional-area ratio $(3/4)m_{\rm{p}}/A$ reduces to $\rho_{\rm p}a$. 
Here, $\rho_{\rm p}$ is the compact particle internal density, $c$ is the gas sound speed, $\rho$ is the gas density, and $\Omega$ is the local disk rotation rate. As this relative flow state takes root, an instability-inducing resonance occurs between the particles and gas 
when the relative stream velocity matches the wave velocity of an inertial wave in the gaseous component \citep{Squire_Hopkins_2018a}.  The strength of the SI is sensitive to the midplane solids-to-gas mass density ratio $\epsilon = \rho_{\rm{solids}}/\rho$, and it is generally considered to be most explosive when $\epsilon \gtrsim 1$ \citep[e.g.,][\citetalias{Umurhan_etal_2020}]{Youdin_Goodman_2005}. 
\par 
In this scenario, the key ingredient for the operation of the SI is   
the normalized radial pressure gradient $\beta$, sometimes referred to as the headwind parameter, given by
\begin{equation}
    \beta = - \frac{1}{2}\left(\frac{c}{v_{\rm{K}}}\right)^2 \frac{\partial \,{\rm{ln}}\,p}{\partial \,{\rm{ln}}\, r},
\end{equation}
where 
$v_{\rm{K}}$ is the local orbital velocity and the pressure $p = \rho c^2$. 
This definition follows from the definition of the $\eta$ parameter found in Eq. (8) of \citet{Cuzzi1993}. 
However, for this work we use the background pressure gradient $\Pi$ defined by the normalized difference between the local orbital velocity and the pressure supported azimuthal gas velocity, $\beta v_{\rm{K}}$ as \citep[e.g.,][]{Carrera_Simon_2022}

\begin{equation}
    \Pi = \frac{\beta v_{\rm{K}}}{c}.
\end{equation}

We define $Z_c$ 
to be the critical metallicity ({\postrevisionbf{or critical solids abundance, e.g. see \citealt{Lesur_etal_2022}}}) above which the SI is predicted to lead to sufficiently large amplitude gravitationally bound overdensities. 
While these critical metallicities are 
under some debate \citep[\citetalias{Umurhan_etal_2020};][]{Carrera_Simon_2022}, $Z_c$ is generally found to be a function of St, $\Pi$ and $\alpha$. The occurrence or absence of strong SI depends on whether or not the particle-gas layer experiences external turbulent forcing or self-generated dusty-gas instabilities. 
\par
Based on analysis of SI simulations without any external turbulent forcing ({\it i.e.} globally laminar yet turbulent due to dusty-gas midplane layer instabilities), \cite{Sekiya_Onishi_2018} conjecture that the relevant parameter determining strong activity is $Z/\Pi$, and its critical marginal value $Z_c$ is only a function of St. 
In this respect there are three proposed occurrence diagrams based on a series of globally laminar axisymmetric and fully 3D simulations.  The first two of these are based on the simulations of \citetalias{Carrera_etal_2015} in which
\beq
\log\left(\frac{Z_c}{\Pi}\right) = 
0.2 \left(\log {\rm{St}}\right)^2 + 0.59 \log {\rm{St}} - 0.27,
\label{equ:car}
\eeq
and \citetalias{Yang_etal_2017}, with
\beq
\log\left(\frac{Z_c}{\Pi}\right) = 
0.1 \left(\log {\rm{St}}\right)^2 + 0.2 \log {\rm{St}} - 0.46,
\label{equ:yang}
\eeq
\citep[see also][]{Carrera_Simon_2022}.  
The third of these is based on \citetalias{Li_Youdin_2021} who report that the range of SI activity in their globally laminar models is greatly expanded to include substantially lower values of $Z$ for $\St > 0.015$.  
They propose that
\beqa
\label{equ:li}
& & \log\left(\frac{Z_c}{\Pi}\right) =  \\
& & \left\{
\begin{array}{cc}
0.1 \left(\log {\rm{St}}\right)^2 + 0.32 \log {\rm{St}} - 0.24; & \ \ \ \St < 0.015; \\
0.13 \left(\log {\rm{St}}\right)^2 + 0.1 \log {\rm{St}} - 1.07; & \ \ \ \St > 0.015
\end{array}
\right. .
\nonumber
\eeqa
\par
The corresponding occurrence or absence of strong SI in the presence of external sources of turbulence is, like the laminar-case critical metallicities above, also subject to debate with no current consensus on the matter \citep[e.g, \citetalias{Umurhan_etal_2020};][]{Chen_Lin_2020,Gole_etal_2020,Schafer_etal_2020}. The uncertainty centers on the character of the underlying turbulence, e.g., where and on what scales is it mainly isotropic or strongly anisotropic \citepalias[see recent mention in][]{Schafer_Johansen_2022}.  
In addition to the occurrence in Eq. (\ref{equ:li}), 
\citetalias{Li_Youdin_2021} (see their section 3.3.2) have also suggested an analogous relationship for $Z_c$ that explicitly takes into account the degree of turbulence present in terms of the $\alpha$ parameter. Symbolically distinguished here as $Z_{c,\alpha}$, this criterion is based on simulations reported both in their study of self-generated SI turbulence and that of \citet{Gole_etal_2020} who considered the formation of overdensities in a model of the SI submerged in an externally driven turbulent fluid.  
It was proposed that
\beq
\label{equ:Zca}
\frac{Z_{c,\alpha}}{\Pi} = \epsilon_c(\St)\sqrt{\frac{1}{25} + \frac{\alpha/\Pi^2}{\alpha + \St}}, 
\eeq
in which it is assumed that the laminar critical solids-to-gas ratio $\epsilon_c$ also applies in the presence of external sources of turbulence:
\beqa
\label{equ:epc}
& & \log \epsilon_c(\St) = \\ 
& & \left\{
\begin{array}{cc}
\log (2.5); & \ \ \ \St < 0.015; \\
0.48 \left(\log {\rm{St}}\right)^2 + 0.87 \log {\rm{St}} - 0.11; & \ \ \ \St > 0.015
\end{array}
\right. . \nonumber
\eeqa

\noindent
In this treatment we see that, for a given St, the effect of including a source of external turbulence is to shift the region of SI activity upwards. However, we note a caveat that this extension of the laminar case neglects some of the physics of turbulence in pp-disks such as radial diffusion which, for instance, is included in the 
solutions of \citetalias{Umurhan_etal_2020}, and which becomes increasingly important for decreasing St.


\par
\citetalias{Schafer_Johansen_2022} also propose $Z-\St$ threshold boundaries for strong acting SI  inferred from global axisymmetric simulations of the VSI in the presence of particles.  We focus on their so-called {\it ``SIafterVSI''} suite of simulation runs in which the VSI is allowed to develop before particles of a selected St value are added into a given numerical experiment. We do this mainly for two reasons: (1) because such a modeled scenario is more realistic as the VSI is likely well developed and turbulent before sub-micron sized grains can grow to the targeted St 
\citepalias[e.g.,][also Sec. \ref{sec:traj}]{Estrada22_paper2}; and, (2) of the various models examined in \citetalias{Schafer_Johansen_2022} those of the {\it SIafterVSI} batch envision a wider area of $Z-\St$ parameter space predicting strong acting SI in the presence of VSI turbulence.  Based on earlier published simulations of similar setup \citep{Schafer_etal_2020} the VSI-SI complex appears to lead to $\alpha = \order{10^{-4}}$ for the $0.005\le Z \le 0.02$ values examined\footnote{We note here that the turbulent intensity of the VSI has been shown by \citet[][and see also \citealt{Lehmann_Lin_2022}]{Lin_2019} to weaken as the particle loading increases.}. We will adopt this value forthwith when considering the VSI.

Equations (\ref{equ:car}-\ref{equ:epc}) define the so called SI occurrence boundaries. 
\citetalias{Schafer_Johansen_2022} do not provide an occurrence fit like the ones discussed previously, and only sketch the boundaries of this active zone to be simultaneously greater than St = 0.01 and $Z=0.0025$ with a straight line connecting the parameter pairs $(\St,Z) = (0.015,0.01), (0.1,0.0025)$ on a semilog graph. The simulations reported in \citetalias{Schafer_Johansen_2022} are characterized by values of the background pressure gradient 
given by 
$\Pi = 0.082\, (r/10\, {\rm{au}} )^{1/4}$. 
We have translated this proposed occurrence boundary and expressed it in terms of $Z/\Pi$ and St 
as part of our analysis in Sec. \ref{sec:results}. 

\section{Global Models}\label{sec:globmod}

In this paper we analyze a subset of pp-disk simulations from the recent work of \citetalias{Estrada22_paper2} (see also \citealt{Estrada22_paper3}). These models use our $1+1$D global nebula evolution code \citep{Estrada_etal_2016} which includes the self-consistent treatment of growth and radial drift of all sizes, accounts for the vertical settling and diffusion of smaller grains, radial diffusion and advection of multiple species in solid or vapor phase, and contains a self-consistent calculation of the opacities and temperature (which depends on the evolving size distribution) that allows one to track the evaporation and condensation of all refractories and volatiles as they are transported throughout the gas disk \citep[see also][]{Sengupta2022}.

\begin{figure}[t]
\begin{center}
\includegraphics[width=0.45\textwidth]{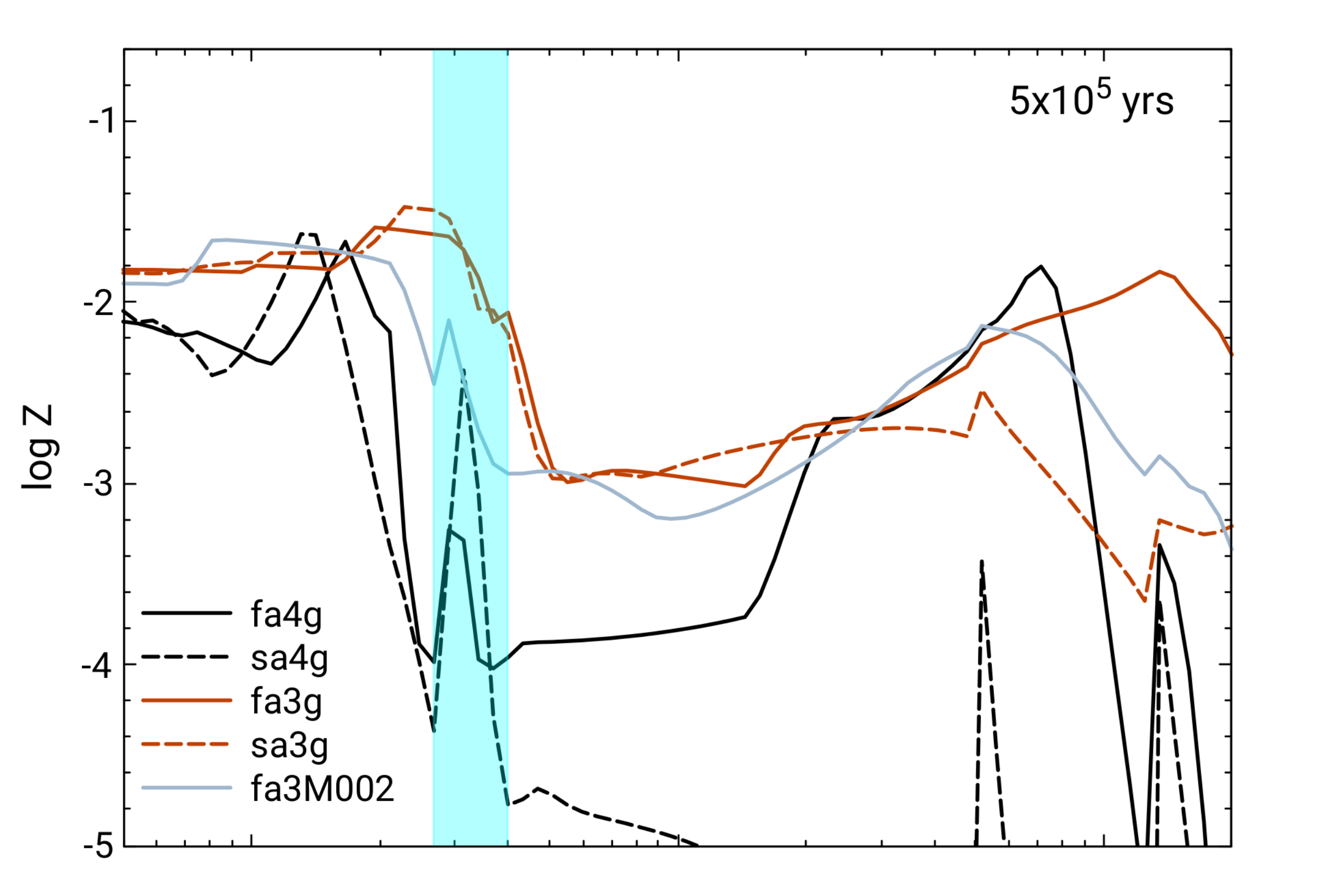}\\
\includegraphics[width=0.45\textwidth]{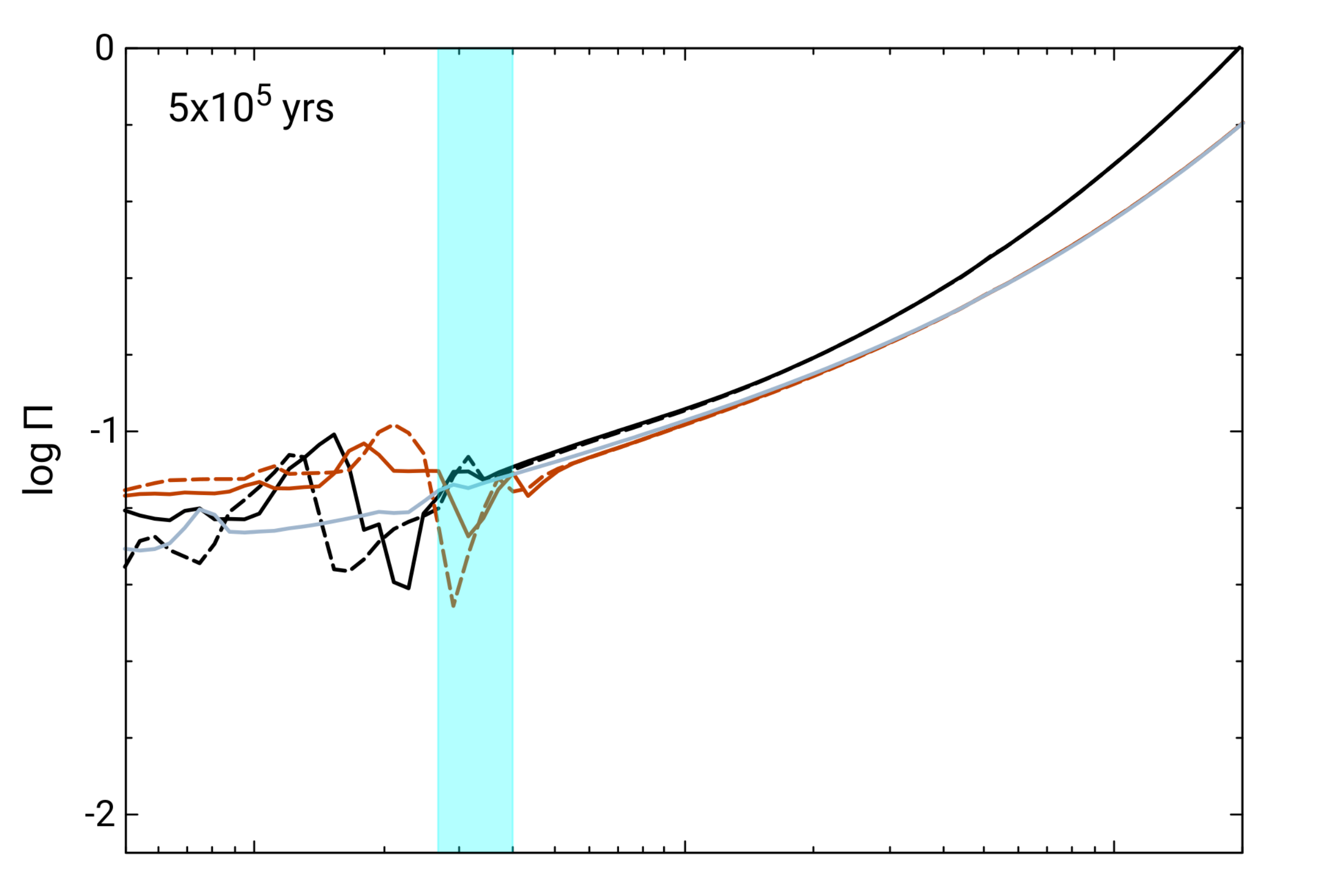}\\
\includegraphics[width=0.45\textwidth]{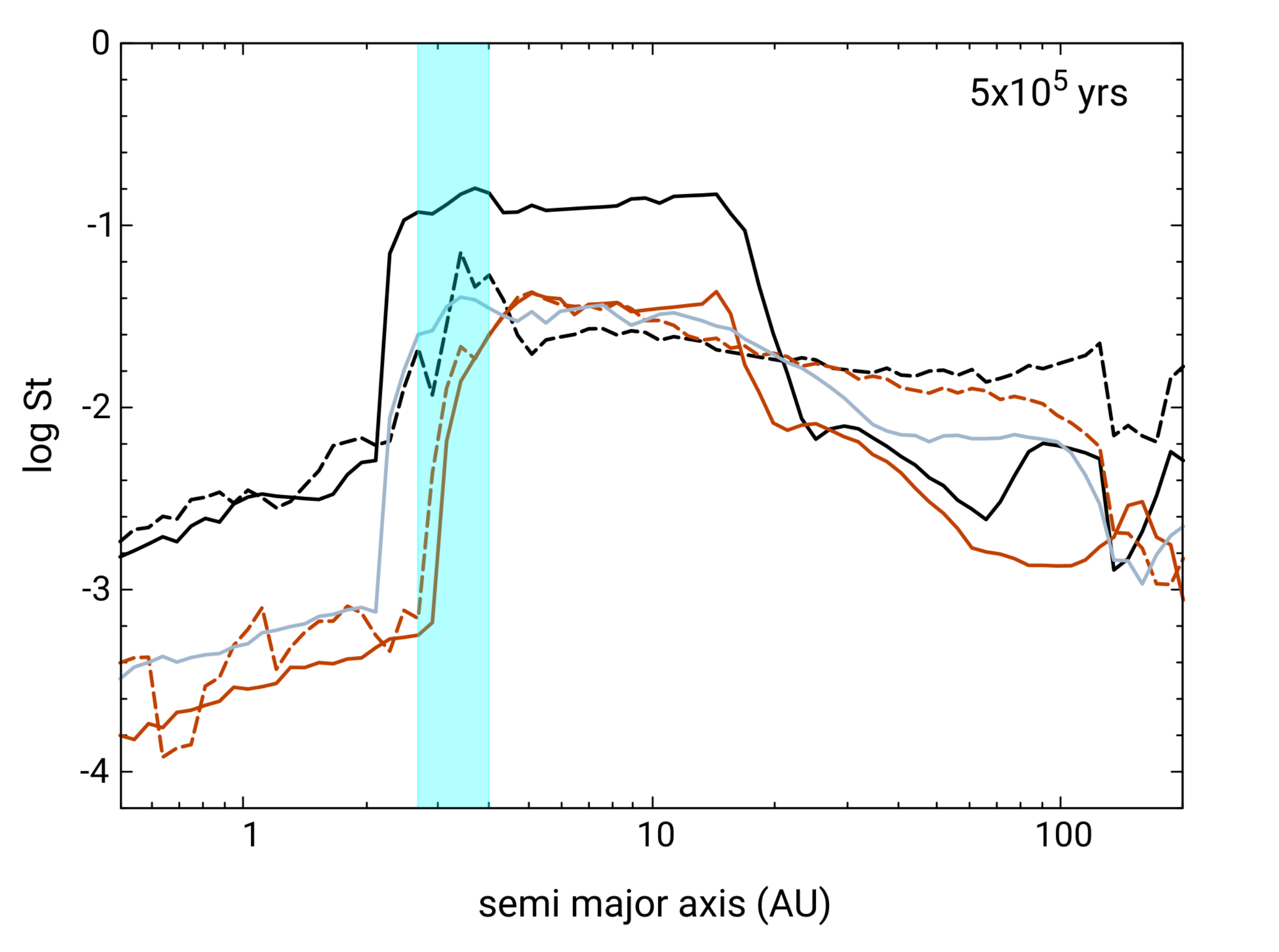}
\par
\end{center}
\vspace{-0.2in}
\caption{Local disk metallicities $Z$, 
normalized pressure gradient $\Pi$, 
and the mass dominant particle Stokes number St for all models as a function of semi-major axis after 0.5 Ma. Fractal growth models are given by solid curves, and compact particle growth models by the dashed curves. The shaded cyan region demarcates the radial range over which the water snowline is located after this time. 
Regions of enhancement in $Z$ 
are generally outside evaporation fronts \citep[see][]{Estrada22_paper3}. One additional lower disk mass model (grey curves) of 
$0.02$ M$_\odot$ (fa3M002g), ten times lower than model fa3g, is included for comparison. 
}
\label{Figure_Collection_1}
\end{figure}

\citetalias{Estrada22_paper2} have used this model to do a comprehensive study comparing compact particle and porous aggregate growth and drift over a range of disk conditions for a 1 M$_\odot$ star, by including a model for fractal aggregate growth and compaction due to collisional \citep{Oku12,Suy12} and non-collisional effects \citep{Kat13a,Kri15}. We select simulations from this work with both compact and fractal aggregate growth for global turbulent intensities of $\alpha=10^{-3}$ and $\alpha=10^{-4}$. These comprise a reasonable range of $\alpha$-values consistent with observational indications 
that suggest that turbulence in pp-disks may be at the level of $10^{-4} < \alpha < 10^{-3}$ \citep[e.g.,][]{Teague2016,Flaherty2017,Flaherty2018}. 
The new aspect here is that these simulations \citepalias[specifically models sa3g, fa3g, sa4g, fa4g; see Table 2 of][]{Estrada22_paper2} have been evolved further from 0.5 Ma to 1 Ma to provide a sufficiently long baseline for our purposes. 

The initial conditions for these models are derived from the analytical expressions of \citet{LP74} as generalized by \citet{Har98} in terms of disk mass 
and turnaround radius 
(initially $\sim 10$ au) where the mass flux changes sign. 
The initial disk mass for these models is 
$0.2$ M$_\odot$ with a total metallicity of $\bar{Z}\simeq 0.014$. 
The initial gas surface density 
is fairly compact similar to that of \citet{Des07}, and the disks are warm, with the temperature 
dependent on viscous dissipation and a time variable stellar luminosity which is roughly an order of magnitude higher than solar at the beginning of the simulations. This places the water snow line initial location between $\sim 10-15$ au and is meant to represent early (Class I) pp-disk conditions immediately following the infall stage \citep[see, e.g.,][]{DD18,HN18}. Detailed discussion of the initial conditions can be found in \citetalias{Estrada22_paper2}.

Figure \ref{Figure_Collection_1} provides a snapshot of these simulations after 0.5 Ma of evolution. We plot the local metallicity $Z$ (top panel), the normalized pressure gradient $\Pi$ (middle panel) 
and the {\it mass dominant} particle and aggregate Stokes numbers (bottom panel) for both fractal aggregate (solid curves) and compact particle (dashed curves) growth models as a function of location in the disk. By mass-dominant, we mean the particle size (mass) defined by the mass weighted mean of the distribution. 
Our simulations have size distributions that range from a 0.1 micron monomers to the largest particle and aggregate size reached before growth frustration due to bouncing, fragmentation, and/or radial drift sets in. Though their St may be similar, the underdense fractal aggregates can be quite large compared to their compact counterparts \citepalias[see Appendix of][]{Estrada22_paper2}.
 
As noted by \citetalias{Estrada22_paper2}, a general difference between fractal and compact particle growth models is that radial drift is less of a factor for porous aggregates over a significant fraction of their growth phase due to a combination of initially pure fractal growth and then slow compaction, while the St of the compact particles increase more quickly and are subject to stronger radial drift earlier on. Eventually fractal aggregates can grow much faster due to their larger cross sections. This behavior generally explains the difference between compact and fractal models in $Z$, especially outside of $\sim 20$ au (Fig. \ref{Figure_Collection_1}, upper panel). The difference in St (lower panel) inside and outside the water snowline (the cyan region across these models between $\sim 2.8-4$ au after this time) is due to water ice being more sticky, which allows water ice-bearing aggregates to grow much larger before becoming fragmentation or drift limited.

It is important to understand the variation in the local metallicity $Z$. Enhancements in the local metallicity (found to reach values as high as $Z\sim 0.05$ in these models) arise primarily due to the radial drift of particles and aggregates across evaporation fronts (EFs). Inward drifting particles can lose most or all of their associated volatile enhancing the vapor phase inside the EF. This vapor can subsequently diffuse (or advect) outwards back across the EF and recondense, enhancing the solids abundance there. In cases of low $\alpha$ and/or at longer times, these enhanced regions can appear as ``bands'' dominated by the associated volatile, most notably for the compact growth model sa4g for $\alpha=10^{-4}$ in the top panel of Fig. \ref{Figure_Collection_1} \citep[see also Fig. 6, upper right panel of][and for more detailed discussion]{Estrada22_paper3}.





\begin{figure*}
\begin{center}
\includegraphics[width=8.65cm]{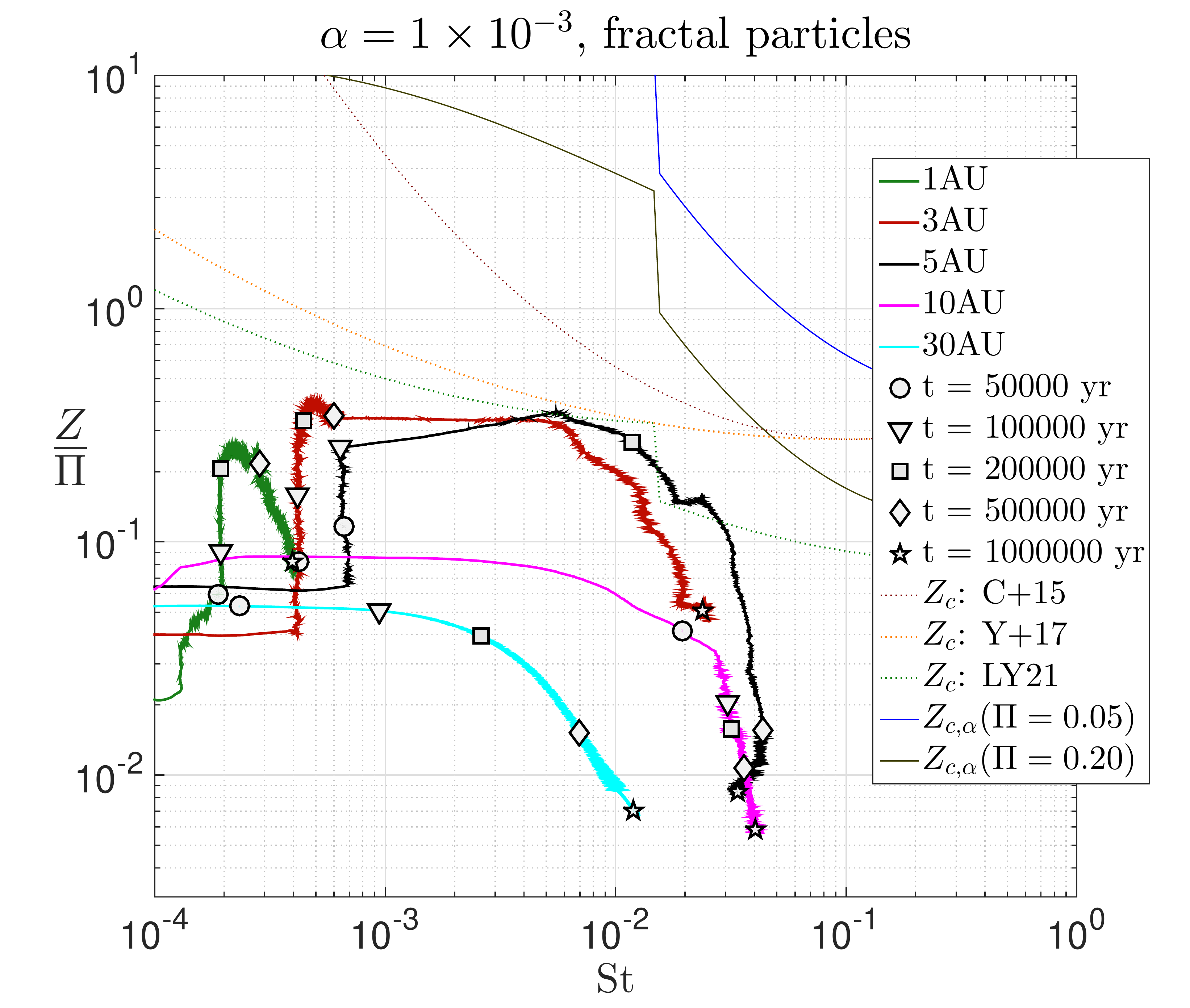}
\includegraphics[width=8.65cm]{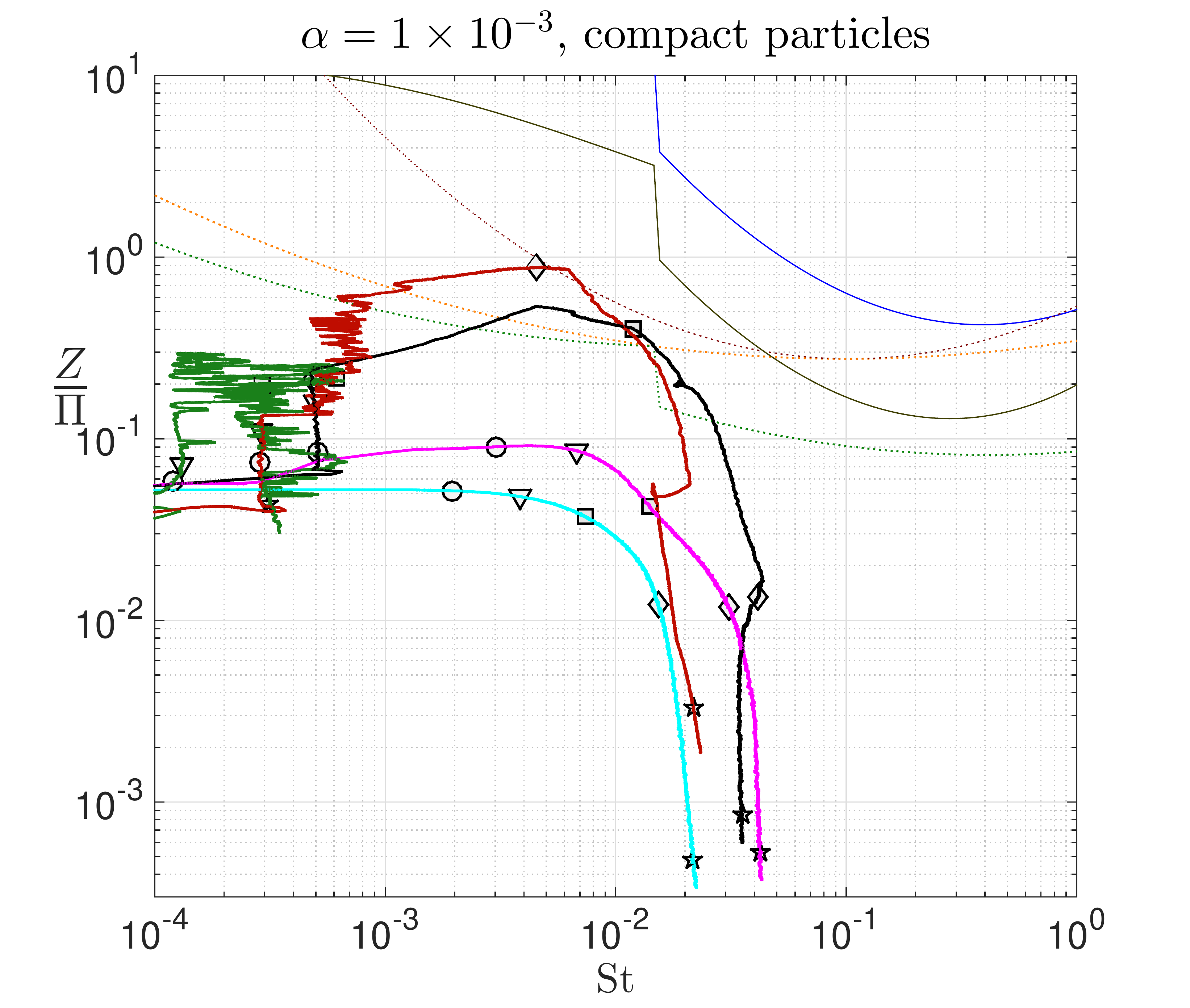}\\
\includegraphics[width=7.85cm]{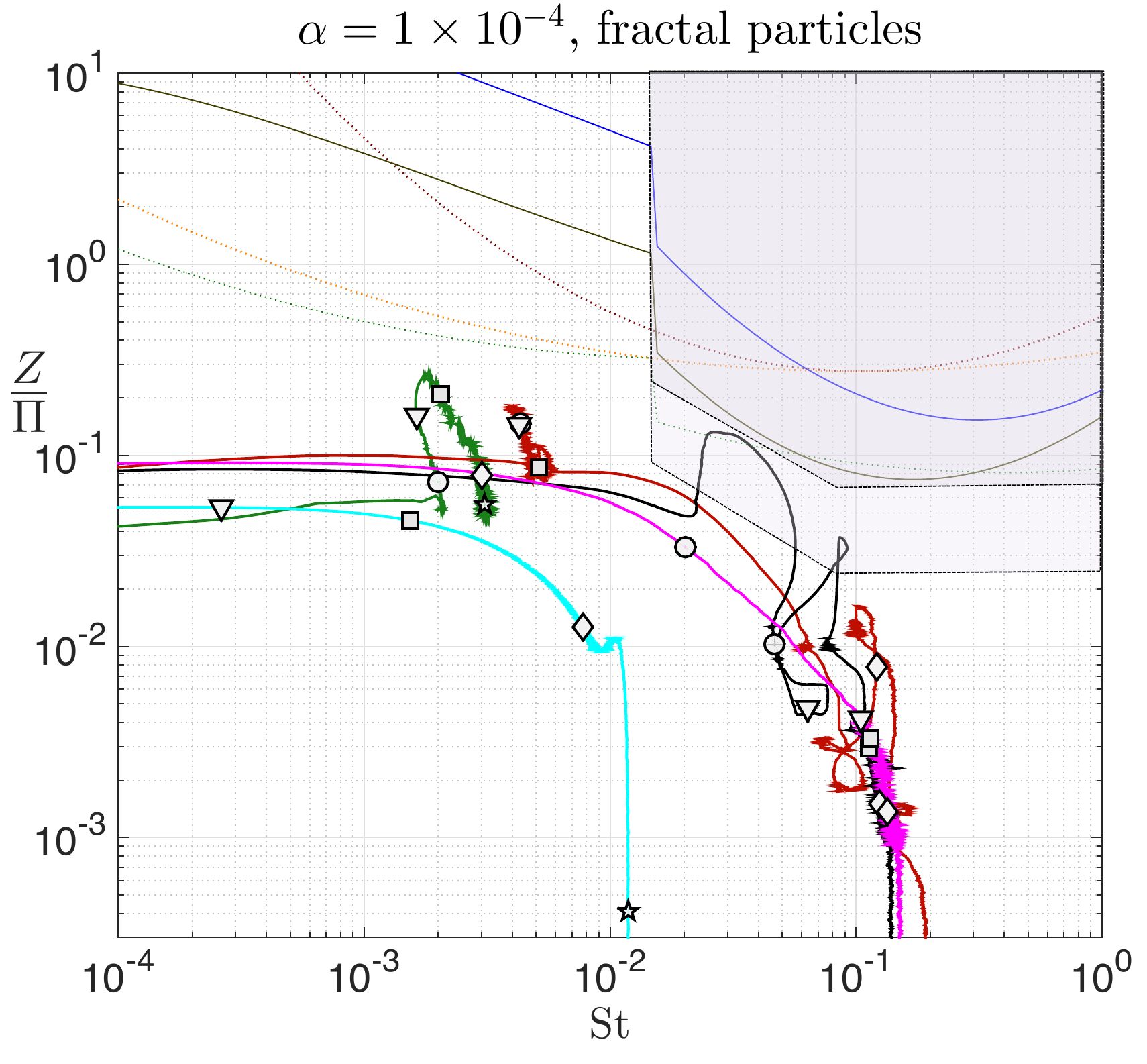}
\hskip 0.7cm
\includegraphics[width=7.85cm]{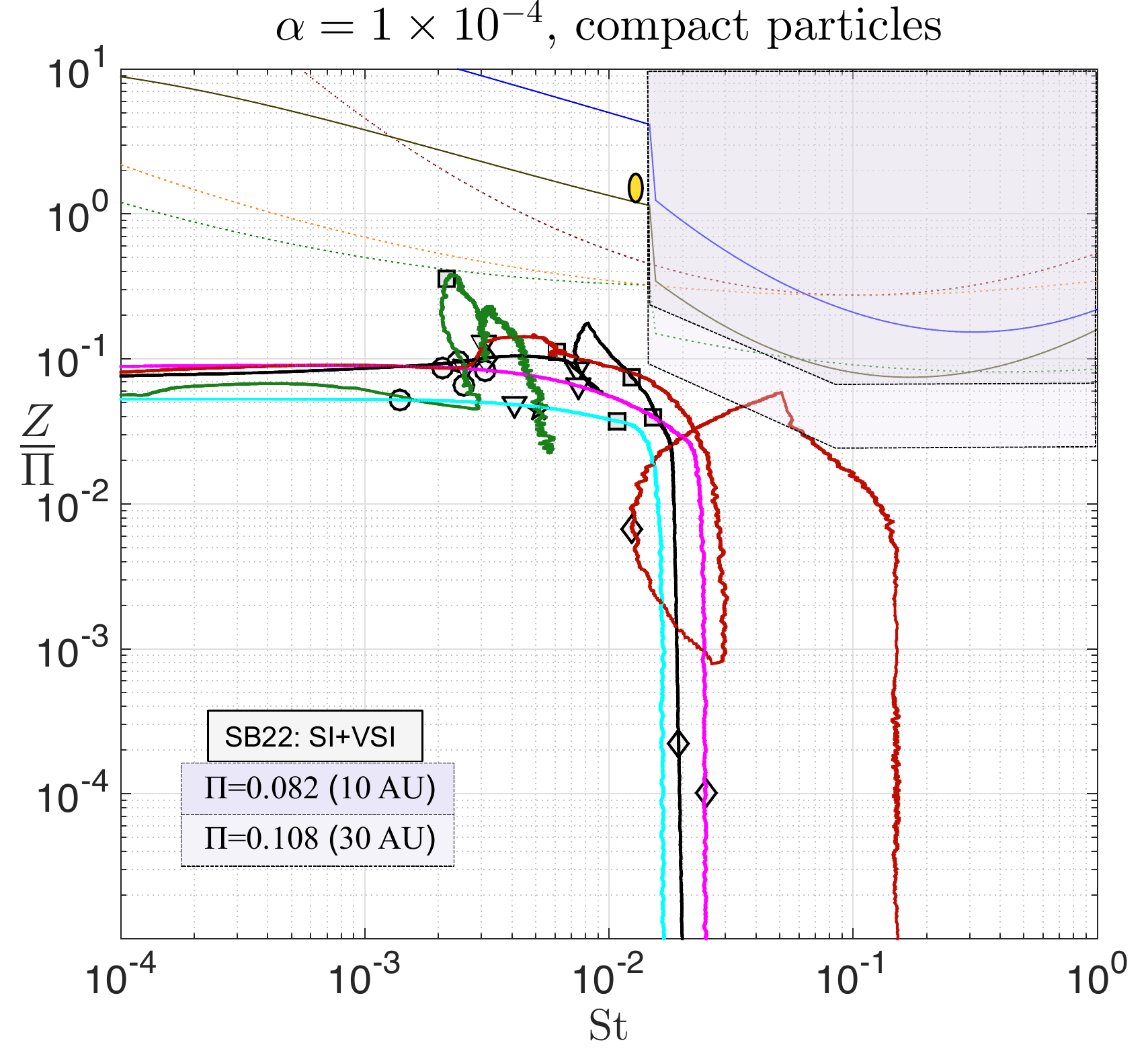}
\par
\end{center}
\caption{Temporal trajectories in $Z/\Pi-\St$ parameter space for the global pp-disk evolution models with compact particle and fractal aggregate growth discussed in Section 3. Trajectories for each model (which begin far off to the left) are plotted at 1, 3, 5, 10, and 30 au spanning 1 Ma for each simulation: (upper left) fractal particles and $\alpha = 10^{-3}$;
(upper right) compact particles and $\alpha = 10^{-3}$;
 (bottom left) fractal particles and $\alpha = 10^{-4}$;
(bottom right) compact particles and $\alpha = 10^{-4}$.  
These correspond to models fa3g, sa3g, fa4g, and sa4g, respectively, in Fig. \ref{Figure_Collection_1}.
Also shown are the predicted critical curves $Z/\Pi$ as a function of St, $\Pi$ and $\alpha$ based on Eqns. (\ref{equ:Zca}-\ref{equ:epc}). In the bottom panels, we additionally show the occurrence region predicted from the simulations of \citetalias{Schafer_Johansen_2022} for two values of $\Pi$ as indicated in the bottom right legend. Open symbols along the trajectories mark the associated times in the upper left legend.
}
\label{Figure_Collection_2}
\end{figure*}

\section{Interpretation of results in terms of SI}\label{sec:results}



 \subsection{Temporal Trajectories of Particle Growth and Disk Evolution}\label{sec:traj}

In Figure \ref{Figure_Collection_2}, we plot the temporal trajectories of particle St and $Z/\Pi$ over 1 Ma for the models introduced in Sec. \ref{sec:globmod}. Each of the trajectories for the indicated radial locations in the disk (1, 3, 5, 10 and 30 au) begin far off to the left on the $x$-axis. A typical monomer in the regions of interest have ${\rm St} \sim 10^{-9}$, though this will depend on the gas surface density. Growth rates vary throughout the disk, but at 1 au for instance where growth rates are more rapid, St can increase by a few orders of magnitude in only 100 orbits\footnote{Note that well developed VSI is found to occur in $< 100$ orbital periods \citep{Richard_etal_2016}.} 
for the compact growth model, while the initially slower growing fractal aggregates have St an order of magnitude smaller than the compact case 
after the same time. 

The sharp increases (or decreases) in $Z/\Pi$ seen along the trajectories can be associated with {\postrevisionbf{multiple refractory species' EFs inside the snowline \citep[e.g., silicates, FeS, see][]{Estrada22_paper3}}} 
which should be understood to be evolving inwards as the disk cools with time. 
Recondensing volatiles from the migrating particles that cross these EFs can lead to enhancements in solids outside them (Sec. \ref{sec:globmod}), stimulating growth to larger St there. 
For example, the large variation in $Z/\Pi$ seen at 3 au in the compact growth case for $\alpha=10^{-4}$ is due to the water snowline evolving through this radius. As discussed in Sec. \ref{sec:globmod}, this particular model (see Fig. \ref{Figure_Collection_1}, upper panel) is characterized by bands 
of enhancements 
which can be orders of magnitude higher than the background $Z$. As the EF migrates inwards, the location of these bands also move inwards explaining the even more sharp decrease in $Z/\Pi$ relative to the other models. 

Interestingly, simulations with the higher $\alpha=10^{-3}$ tend to have larger $Z/\Pi$ than the lower $\alpha$ case and which persist for longer periods of time. 
The inward migration of the EFs from the water snowline and inward is slower for $\alpha = 10^{-3}$ because for this level of turbulence the inner disk stays hotter longer due to combination of particle size and opacity \citep[e.g., see Fig. 1,][]{Estrada22_paper3}. As a result this can lead to larger local metallicity. 
For $\alpha=10^{-4}$, cooling is more rapid because the mass dominant aggregates grow quickly to higher St leading to lower opacities, 
and faster drift, so enhancements do not get quite as large. 

Overall, the general trend across these models is that as particles and aggregates grow and their St increases, radial drift becomes faster leading to a sharp, concurrent decrease in the metallicity near large St regions so that having both large St and $Z$ for $\alpha \ge 10^{-4}$ is prevented. 
Thus for this epoch, $\epsilon < 1$, always. {\postrevisionbf{This behavior in our simulations differs from other published work that in some cases find that conditions for the SI can be satisfied ({\it i.e.}, $\epsilon = \order{1}$) for these turbulent intensities. We summarize some key differences between our models and others in the Appendix.}}

 \vspace{0.1in}
 
\subsection{Turbulence and SI}

Also plotted in Fig. \ref{Figure_Collection_2} are the SI occurrence boundaries previously advocated by others for the critical $Z_c$ (dashed curves) defined by Eqs. (\ref{equ:car}-\ref{equ:li}), 
and for $Z_{c,\alpha}$ (solid curves) defined by Eqns. (\ref{equ:Zca}-\ref{equ:epc}). 
Our particle growth simulations tended to achieve $\Pi$ values lying
within a well-defined range $0.05 < \Pi(r,t) < 0.22$ (corresponding to the inner disk region, and $\sim 100$ au, respectively).  As such, we always depict two $Z_{c,\alpha}$ curves corresponding to these two extreme values of $\Pi$. 
All curves for $Z_{c,\alpha}$ and $Z_c$ are specifically labeled in the legend in the top left panel. We see that although there are instances where curves for $Z_c$ are crossed (e.g., top right panel), those crossed apply only to laminar disks (in which the only turbulence is self-generated by the particle layer). We find quite generally that there are no cases where the temporal trajectories of particle growth in turbulence, in either fractal or compact cases, cross the critical curves $Z_{c,\alpha}$ which have been proposed for globally turbulent pp-disks like those we examine here. 

The boundaries in parameter space defining the occurrence or absence of strong acting SI under turbulent conditions remain unsettled.  The three studies \citepalias{Carrera_etal_2015,Yang_etal_2017,Li_Youdin_2021} described in Sec. \ref{sec:SIoccur} established the occurrence boundaries in $Z,\Pi$ and St by analyzing direct numerical simulations of the SI in globally non-turbulent nebulae and querying whether a given numerical experiment generates particle overdensities that exceed the Roche-density condition for gravitational binding. If these requisite conditions are met, then the SI is considered to be strong enough to lead to planetesimals. But these criteria only apply when there is no external source of turbulence. 

We do note that even these proposed criteria appear to be violated in some instances like in the recent study of \citet{Carrera_Simon_2022}, who considered the SI in a model with low St ($=0.015$, 1 mm-sized compact grains) in a globally non-turbulent disk, but experiencing settled particle layer generated turbulence with $\alpha \approx 6\times 10^{-5}$. They found that no sufficiently large gravitational overdensities were produced in the (settled layer driven) turbulently active part of the simulation, 
where locally $Z/\Pi \approx 1.06$ and $0.028 < \Pi < 0.030$.  
The lower row (right panel) of Fig. \ref{Figure_Collection_2} designates the position in parameter space of this simulation with a vertically elongated orange oval that lies well above the appropriate 
criteria $Z_c$ based on purely self-generated turbulence,
Eqns. (\ref{equ:car}-\ref{equ:li}, dashed lines), suggesting that strong overdensities should occur. 
Yet, according to the corresponding externally-driven turbulence occurrence prediction of \citetalias{Li_Youdin_2021}, Eqns. (\ref{equ:Zca}-\ref{equ:epc}, solid lines), this simulation is not expected to produce gravitationally bound overdensities.  \citet{Carrera_Simon_2022} offer a different interpretation for this outcome.  
{\postrevisionbf{According to their {\it {residence time}} idea, the SI might well be on its way toward producing gravitationally bound overdensities in the particle field but that there is not enough time for it to fully develop before the particles radially drift into that part of the disk where the SI ceases to operate. }}
\par
One should also consider a corresponding weak criterion based on SI growth rate estimates in the face of turbulence like the $\alpha$-disk turbulence models of \citetalias{Umurhan_etal_2020} and \citet{Chen_Lin_2020}, where it is envisioned that SI derived high density filaments actually grow out of a largely isotropic turbulent state driven by some process other than the SI itself -- a proposition supported by recent re-examination of particle-laden turbulence in pp-disk midplanes \citep[see][]{Sengupta_Umurhan2023}. 
We propose this to be a weak bookend criterion simply on the fact that not all SI unstable settings are expected to lead to gravitationally bound overdensities.
As the thinking goes, if the isotropic turbulence is so strong that predicted SI growth rates are thousands of orbit times or longer, then it would be justifiable to conjecture that the SI is either very weak or practically inactive.  
\citetalias{Umurhan_etal_2020} in fact show for the parameter combination St $\lesssim 0.1$ and $\alpha$ large enough that the midplane values of $\epsilon \equiv \rho_{\rm solids}/\rho \le 1$ ({\it i.e.}, the so-called Zone II region of the $\alpha - \St$ parameter plane) that the predicted growth rates are not only very long, but also (of the few simulations reported in the literature that were conducted in that parameter range) indeed show no appreciable action driven by the SI for the plausible and simple reason that growth timescales exceed drift loss times.

\par
The framework laid out in \citetalias{Umurhan_etal_2020}, suggests that the SI should be strong when the growth times are short compared to loss times -- that is to say, acting on $\order{1}$ orbital timescales --  consistent 
with the simulations that lead to the emergence of strong high density filaments
as reported by \citetalias{Carrera_etal_2015} and \citetalias{Yang_etal_2017}.  As shown in \citetalias{Umurhan_etal_2020}, as well as in \citetalias{Estrada22_paper2}, none of the detailed particle evolution (growth + drift) models run for values of $10^{-4} \le \alpha \le 10^{-3}$ ever generate parameter combinations that predict turbulent SI growth rates less than several hundreds to thousands of orbit timescales.  This general weakness in predicted turbulent SI growth rates is due to the fact that the simulations never produce mass-bearing particles with St values much exceeding $\sim 0.03-0.04$.
For the models presented here we {\it do} find instances where 
${\rm St} \sim 0.1$; however, as St increases, the local values of $Z$ concurrently decrease to extremely low values ($\ll 0.01$).  The turbulent SI growth rates under these conditions are also extremely long according to the theoretical framework of \citetalias{Umurhan_etal_2020} and \citet{Chen_Lin_2020} except for when St begins to approach and/or exceed 0.1.  

\par
Moreover, the particles themselves are also likely coming up against the radial drift barrier (and indeed, this is why local $Z$ decreases more quickly) in which predicted growth timescales are the same order of magnitude or longer than the radial drift timescales across the disk -- 
{\postrevisionbf{an argument in the same spirit as \citet{Carrera_Simon_2022}'s residence time interpretation.}}

\begin{figure}[t]
\begin{center}
\includegraphics[width=0.45\textwidth]{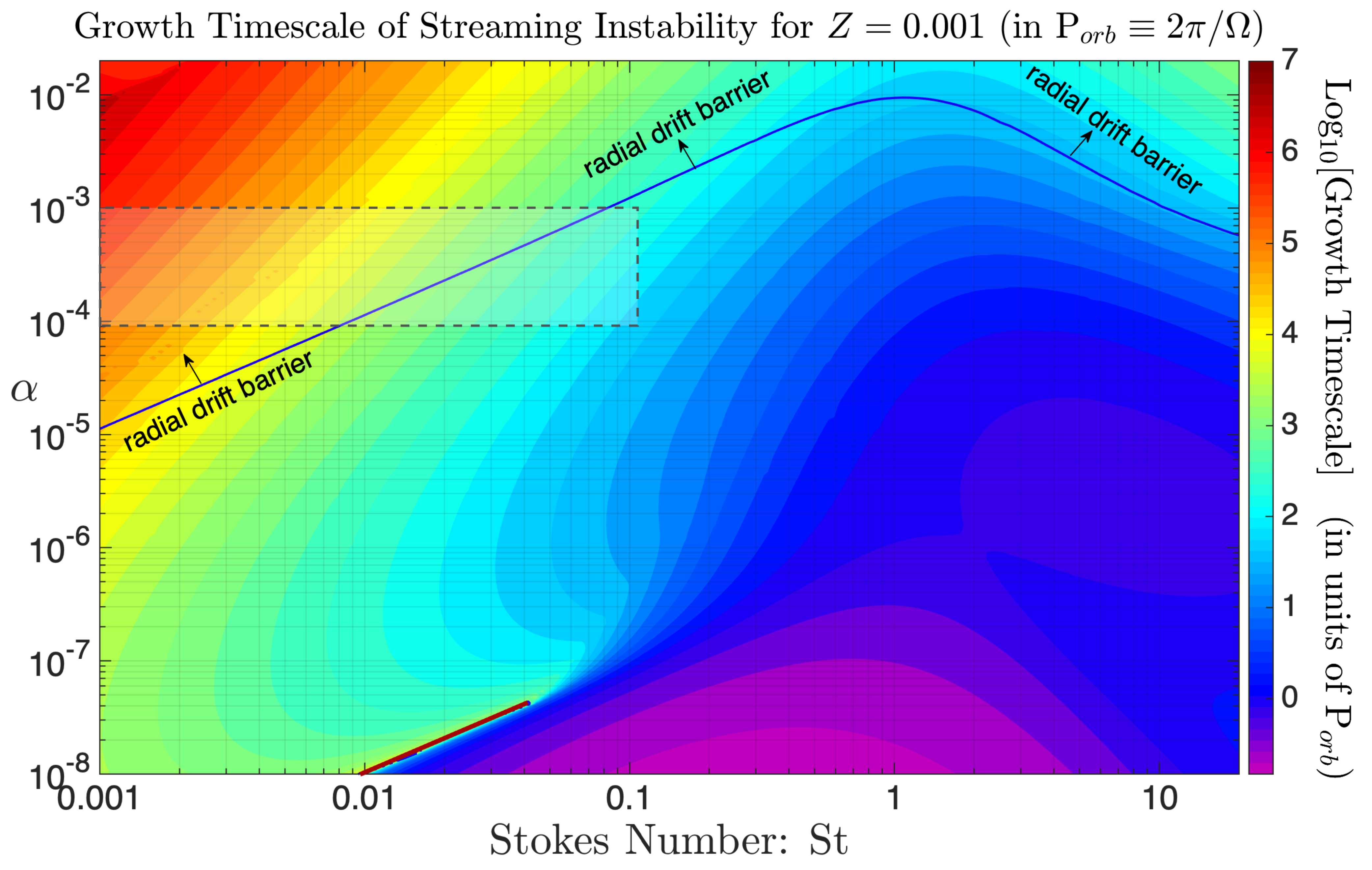}
\par
\end{center}
\vspace{-0.2in}
\caption{SI growth timescales ($t_{\rm g}$) as a function of $\alpha$ and St, with $Z=0.001$ and $\Pi = 0.05$, based on \citetalias{Umurhan_etal_2020} under the assumption of isotropic turbulence.  The hatched box region in upper left demarcates the parameter region generally encompassing those achieved in the simulations reported here.  The figure shows a curve corresponding to the region of parameter space in which $t_{\rm d} = t_{\rm g}$ and designates with arrows where $t_{\rm d} < t_{\rm g}$, where $t_{\rm d}$ is the drift time.  There exists a putative range within the hatched region in which $t_{\rm d}>t_{\rm g}$, but even here growth times are long (see text for further details).}
\label{Figure_Collection_3}
\end{figure}

To examine this competition, we show in Figure \ref{Figure_Collection_3} the \citetalias{Umurhan_etal_2020} predicted growth timescales, $t_{\rm g}$, for the case where $Z=0.001$.  
{\postrevisionbf{ This metallicity value is chosen since it corresponds to an upper $Z$ bound of our simulations during late times -- roughly after 0.2 Ma -- and that which corresponds to largest St values achieved, i.e., around $\order{0.1}$ for models fa4g and sa4g.}}
Overlain on Figure \ref{Figure_Collection_3} are the timescale limits for radial drift, $t_{\rm d}$, based on Eq. (25) of \citetalias{Umurhan_etal_2020}, and reproduced here in terms of orbital time units $P_{{\rm orb}} \equiv 2\pi/\Omega$:
\beq
t_{\rm d} = 
\frac{(1+\epsilon)^2 + {\rm St}^2}{4\pi {\St} \cdot  \Pi^2} P_{{\rm orb}},
\qquad
\epsilon \approx Z\sqrt{\frac{\alpha + {\St}}{\alpha}},
\eeq
in which the relationship for $\epsilon$ is a robust estimate based on the balance of downward vertical particle settling and upward turbulent particle diffusion 
\citep[e.g.][]{Dubrulle_etal_1995}.  
Figure \ref{Figure_Collection_3} also designates the range of $\alpha$ and St values broadly characterizing the particle growth simulations reported here.  
We find that for values of $\St \lesssim 0.01$ neither the low or high $\alpha$ values should lead to linearly unstable turbulent SI growing on timescales shorter than $t_{\rm d}$.  However in the range $0.01 \lesssim \St \le 0.1$ there appears a region in the $\alpha-\St$ parameter plane, which is bounded below by the line $\alpha = 10^{-4}$ and above by the relationship $\alpha \approx 0.01\, \St $, in which $t_{\rm g} < t_{\rm d}$.  In this patch of allowable parameter space we find that $t_{\rm g} \ge 350 P_{{\rm orb}}$, wherein equality occurs for $\alpha = 10^{-4}$ and $\St \approx 0.1$.  We suggest therefore that under these $Z$-rarefied conditions ({\postrevisionbf{which occur for a brief period in the models for $\alpha = 10^{-4}$ when $Z/\Pi \gtrsim 10^{-2}$, see 3 au trajectories in Fig. \ref{Figure_Collection_2}}}) 
that the SI might possibly lead to relatively high density particle filaments inside this parameter region, but we suspect that because of the relatively slow growth rates it is not expected to lead to gravitationally bound overdensities. 
\par
Lastly, we have included the proposed occurrence boundary predicted by \citetalias{Schafer_Johansen_2022} from their {\it SIafterVSI} simulations translated to $Z/\Pi - \St$ space for two values of their assumed $\Pi$ at 10 and 30 au. The occurrence zone is shown as two shaded regions found on the bottom panels of Fig. \ref{Figure_Collection_2} for the cases with $\alpha= 10^{-4}$ which is the most appropriate comparison. The first value corresponds to the inner boundary of their simulations, while the second is used as a direct comparison with our 30 au trajectory. 
It can be seen that although these represent a marked improvement for the favorability of the SI in the presence of a source of external turbulence, the boundaries still lie well above their respective temporal trajectories for 10 (magenta curves) and 30 au (cyan curves) which are already steeply decreasing in $Z$ for $\St \gtrsim 0.01$.  
\par
3D numerical simulations of the VSI conducted on large radial domains predict that its unsteady/turbulent dynamics are strongly anisotropic \citep[e.g.][]{Stoll_Kley_Picogna_2017,Flock_etal_2017,Flock_etal_2020}, with unsteady vertical motions dominating corresponding radial motions.  
\citetalias{Schafer_Johansen_2022} observe that the anisotropic character of VSI turbulence, and its inherently weak attendant radial momentum and particle diffusions, helps to rationalize why the SI can lead to gravitationally bound overdensities in VSI turbulent disks even with moderate turbulence levels ($\alpha  \sim 10^{-4}$) and relatively small effective St (in the range of 0.006 and 0.06). 
By contrast, the SI subject to isotropic turbulence -- yet under otherwise similar St and $\alpha$ measures -- is predicted to be weakened by comparison \citep{Chen_Lin_2020, Umurhan_etal_2020,Gole_etal_2020}.  We understand this to be because the radial diffusion of particles and momentum are of similar order magnitude as that in the vertical direction, which implies that radial diffusion should act against the ready formation of filaments\footnote{\citetalias{Umurhan_etal_2020} show that the fastest growing modes generally correspond to relatively short radial wavelength disturbances compared to the vertical, especially under moderate values of turbulent intensity and values of St $<0.1$.}.

\par

Indeed, a preliminary spectral kinetic energy analysis of the VSI in the radially relatively large domain, fully 3D, numerical experiment reported in \citet{Flock_etal_2020} shows tell-tale signs of Kraichnan-Batchelor dual cascade in the gaseous kinetic energy, whose characteristic inverse energy cascade likely goes in part toward explaining the observed strong anisotropy in well-developed VSI simulations -- especially with respect to its predominant vertical motions -- as well as the emergence of prominent large-scale $z$-direction oriented vortices in sufficiently global disk models. \citep[similarly in VSI simulations of ][]{Manger_etal_2021}
 \par
 {\postrevisionbf{
It is worth commenting that \citet{Flock_etal_2020} show the kinetic energy distribution (${\cal E}_m$) on the spectrum of azimuthal wavenumbers ($m$) within a restricted annular regime of their full simulation. They find a ${\cal E}_m \sim m^{-3}$ dependence on lengthscales shortward of a characteristic scale $\lambda_0 \equiv 2\pi R/m_0$, and a ${\cal E}_m \sim m^{-5/3}$ shape longward of $\lambda_0$.  This feature is emblematic of Kraichnan-Batchelor double cascade character of 2D turbulence driven at a length scale $\lambda_0$ \citep{Boffeta_Ecke_2012, Alexakis_Biferale_2018}, where energy propagates toward scales larger than $\lambda_0$, while enstrophy propagates toward scales smaller than $\lambda_0$. 
Close inspection of the simulation discussed  would show
that the lengthscale $\lambda_0$ corresponds to the emergence of radially/azimuthally shortened but vertically elongated columnar vortices \citep[e.g., see Fig. 5 in][]{Flock_etal_2020}. 
In accordance with an inverse-energy cascade dynamic,
 we think that given enough time these vortices  merge with one another to ultimately bring about the aforementioned azimuthally extended large-scale steady vortices seemingly characteristic of well-developed VSI.}}
 \par
 {\postrevisionbf{
 Whether the VSI is truly quasi-2D in dynamical character or, perhaps, dynamically more akin to flows exhibiting a split-cascade remains to be determined only once higher resolution simulations are conducted that capture its fastest growing linear modes with length-scale $\lambda_{{\rm max}}$.  For the VSI $\lambda_{{\rm max}} = \order{H^2/R}$  \citep{urpin03,Umurhan_etal_2016,Yellin-Bergovoy_etal_2021}, which needs at least 20-40 grid point resolution across $\lambda_{{\rm max}}$ to numerically resolve any potential downscale energy cascade \citep[for instance, see the relative high resolution but radially restricted VSI simulations of][]{Richard_etal_2016}.  
 }}

On the other hand, if the disk is subject to ZVI turbulence \citep[Zombie Vortex Instability,][]{Marcus_etal_2013, Barranco_etal_2018}, then its turbulence is known to be more isotropic \citep[e.g.][]{Marcus_etal_2016} expressing a classical Kolmogorov $k^{-5/3}$ spectrum down to the smallest dynamically resolvable {\it and} non-artificially dissipating scales.  
Under such driving, using an isotropic turbulence model -- like the simple mixing-length $\alpha$-disk form often used -- would be more appropriate to predict how the SI develops under turbulent forcing.  
Thus, assessing how the SI fares under realistic and properly resolved numerical models of disk turbulence, whether it be driven by the VSI or something else, remains an open issue needing  elucidation in the near term.



\section{Summary}\label{sec:summary}

We analyze previously conducted self-consistent particle growth simulations of \citetalias{Estrada22_paper2} for both fractal and compact aggregates in the presence of global turbulence with values of $\alpha = 10^{-4}-10^{-3}$ in order to evaluate whether conditions for the SI to produce gravitationally bound particle overdensities can be achieved in the earliest stages ($\lesssim 1$ Ma) of globally evolving pp-disks, an epoch in which meteoritic and observational evidence strongly suggest that planetesimals have already formed.  For this work, the selected simulations have been extended from 0.5 Ma to 1 Ma in order to provide a long enough baseline such that the likelihood for strong and efficient acting SI to arise is systematically moving in the direction of lower probability after these times. 

As input, we utilize the time series generated during each simulation 
that track the change in the relevant properties, namely the local metallicity $Z$, the mass-dominant particle or aggregate St, and the pressure gradient as a function of semi-major axis in the disk. 
We then compare temporal trajectories {\postrevisionbf{for several radial locations of interest}} of the constructed $Z/\Pi$ from these 
with the SI occurrence boundaries in $Z$, $\Pi$ and St (Eqns. \ref{equ:car}-\ref{equ:epc}) from four recent studies \citepalias{Carrera_etal_2015,Yang_etal_2017,Li_Youdin_2021,Schafer_Johansen_2022} established via analysis of numerical simulations of the SI. The bulk of these occurrence boundaries only apply to globally laminar disks; however, \citetalias{Li_Youdin_2021} also proposed an extension of this formalism, $Z_{c,\alpha}$, envisioned to apply to a globally turbulent pp-disk, while \citetalias{Schafer_Johansen_2022} also predicted an occurrence boundary based on their simulations of the VSI, both of which we have used here to compare with our models of turbulent particle growth.

\begin{figure}[t]
\begin{center}
\includegraphics[width=0.45\textwidth]{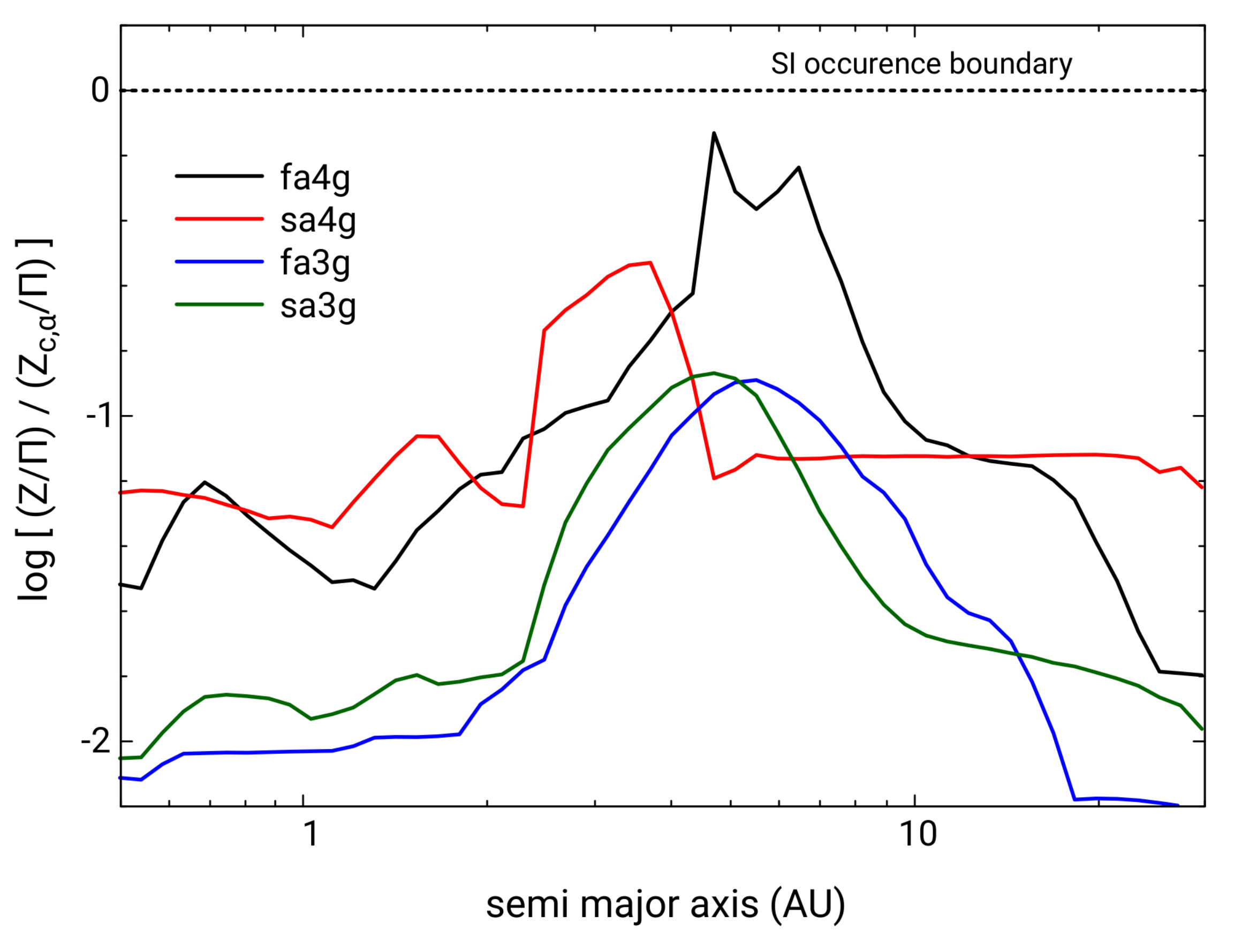}
\par
\end{center}
\vspace{-0.2in}
\caption{{\postrevisionbf{Maximum value for the ratio of $Z/\Pi$ and $Z_{c,\alpha}/\Pi$ from 0.5 to 30 au over 1 Ma for all models considered. To generate these curves, the models' evolving $\St(t)$ and $\Pi(t)$ at each radial location are input into Eq. \ref{equ:Zca} and compared with the models' $Z/\Pi$ over all times $t$ to find the maximum. Each radial location reaches its maximum value at different times and has a different St associated with it. Crossing the SI occurrence boundary (black dotted curve) requires a ratio larger than unity. The largest value occurs in the fractal aggregate model (fa4g) for $\alpha = 10^{-4}$ near 5 au ({\it cf.} Fig. \ref{Figure_Collection_2}) between 30 and 40 kyr. The trend outside of 30 au continues to be the same because $Z/\Pi$ decreases, and $\St$ remain small \citep[though photoevaporation may change these trends at later times, e.g.,][]{Carrera2017}}}}  
\label{Figure_Collection_4}
\end{figure}
 
We find quite generally within the first Ma of evolution that there are no times for any of the analyzed models where the temporal trajectories {\postrevisionbf{at the selected radial locations}} in $Z/\Pi - \St$ space comes near exceeding $Z_{c,\alpha}(\Pi,\St,\alpha)$,  which defines the critical solids abundance for planetesimal formation by the SI proposed by \citetalias{Li_Youdin_2021} in the presence of external turbulence. Furthermore, the temporal trajectories for the compact and fractal growth models with $\alpha = 10^{-4}$ also do not come near the boundaries predicted by \citetalias{Schafer_Johansen_2022} in the presence of the VSI despite conditions being more favorable for the SI compared to \citetalias{Li_Youdin_2021}. The main reason for this is that as growth proceeds to larger St values, particles are subject to rapid radial drift, and the local metallicity concurrently decreases sharply to small values ($Z \ll 0.01$), veering trajectories away from the SI occurrence regions.

This result, which applies to both the compact particle and fractal aggregate growth models, is consistent with the predictions of \citetalias{Umurhan_etal_2020}, and \citetalias{Estrada22_paper2} where the straightforward $\epsilon \gtrsim 1$ constraint is applied. 
{\postrevisionbf{Though we have selected specific radial locations in Fig. \ref{Figure_Collection_2} to demonstrate the temporal trajectories of particle growth and local metallicity explicitly, other radial locations produce similar results, as is summarized in Figure \ref{Figure_Collection_4} where for each radial location we plot the maximum value of each model's $Z/\Pi$-to-$Z_{c,\alpha}/\Pi$ ratio (Eq. \ref{equ:Zca}) achieved over the full 1 Ma course of a simulation. In this format it is easier to visualize that in order to breach the SI occurrence region, this ratio must exceed unity which does not occur for any of our models analyzed here (see figure caption).}}

There are instances where the particle $\St \gtrsim 0.1$, {\postrevisionbf{which are associated with a small local $Z$ owing to fast radial drift}}. \citetalias{Umurhan_etal_2020} only briefly discussed the growth rates for the SI in the context of their theory for a low value of $Z = 0.001$, and here we find that for this $Z$ the growth rates (Fig. \ref{Figure_Collection_3}) are indeed long except when one approaches $\St \sim 0.1$ for $\alpha = 10^{-4}$, in which the e-folding timescales are on the order of several hundred orbits and shorter than the radial drift timescales of the particles.  It is thus feasible that the SI could be active in this epoch under these conditions, though presumably only weakly so, based on the long growth times. 
Whether such conditions could generate planetesimals for these parameters over such long growth times requires addressing with appropriately constructed future numerical simulations.

Finally, we should acknowledge that our results here for the models we have considered do not preclude the operation of the SI at later times when the disk has evolved significantly more due to, for example, open-system loss processes such as winds 
\citep[see, e.g.,][]{Sengupta2022}. For instance, removal of the disk gas via photoevaporation \citep[e.g.,][]{Carrera2017} would lead to an environment where increasing particle St as a result of decreasing gas density may achieve the $Z-\St$ conditions for planetesimal formation via SI, perhaps explaining the formation of KBOs \citep{Nesvorny_etal_2019}, but these would form much later than the epoch addressed in this work. {\postrevisionbf{Along the same lines, low gas mass, very cold disk conditions are more conducive to strong SI, but like the photoevaporative case above these conditions may most often be associated with older, Class II, evolved disks which is not our focus.}}


\begin{acknowledgements}{{\bf Acknowledgements.}}
The authors acknowledge fruitful discussions with the TCAN collaboration including W. Lyra, J. Simon, C.-C. Yang, and A. Youdin, as well as D. Carrera and S. Bonnet. We also thank J. Cuzzi and D. Sengupta for their valuable input, and to an anonymous reviewer whose suggestions helped to improve this paper. We are grateful for the ``Planets in the Desert'' workshop during June 2022 at New Mexico State University that set the scene for us to conduct this study.

\end{acknowledgements}

\vspace{0.2in}
\section*{Appendix \\
Comparison to published results}
\label{sec:appx}

{\postrevisionbf{A reviewer asked that we stipulate the main ways in which 
our models differ from others in the literature 
beyond just model parameter choices. 
We do not attempt an exhaustive comparison here and, instead, consider several studies published over the last decade.
In many cases differences can be distilled down to others' use of simplified, monodisperse growth models in a globally turbulent disk where the {\it largest} particle growth time is taken to be $\tau_{\rm{grow}} = a/\dot{a} \simeq 1/(Z\Omega)$, and in which the SI is {\it assumed} to operate whenever $\St \ge 0.01$ and $\epsilon = \order{1}$ \citep[e.g.,][]{Draz2016,Draz_Alibert2017,Carrera2017,Schoonenberg_etal_2018,Charnoz_etal_2019,Morbidelli_etal_2022}. 
Once these SI criterion are met, it is then assumed that planetesimals form at a somewhat arbitrary rate.

The 
parameterization for $\tau_{\rm{grow}}$ cited above leads to overall much faster growth than in our models mainly because it assumes perfect sticking up to the fragmentation barrier (and/or drift barrier). Moreover, in deriving the approximation $\tau_{\rm{grow}}$ it is also assumed that particles are always in the Epstein regime (all simulations analyzed here have a fraction, and in some cases sizeable, of particles in the Stokes regime, see \citetalias{Estrada22_paper2}), and that $\St \gg \alpha$ even in the initial growth stages. These approximations mean $\tau_{\rm{grow}}$ itself is always independent of particle size (or St). Though a useful tool, these simplified growth models cannot fully capture the complexities that can arise using more detailed coagulation models, and indeed it is known that they deviate from them \citep[e.g., see][]{Lenz_etal_2020}.
\par
Other models largely ignore particle bouncing as well, even though \citet[][also, \citealt{Guttler_etal_2010}]{Zsom_etal_2010} showed that it is a significant barrier to growth. In our model, we do not assume instantaneous fragmentation at some threshold relative velocity, but rather treat fragmentation and bouncing together using a mass- and velocity-dependent sticking coefficient \citep[e.g., see also][]{Okuzumi_Hirose2012,Homma_etal_2019,Estrada_etal_2016}, which can increasingly slow the growth rate as the fragmentation barrier is approached because the largest particle can only grow efficiently from a relatively diminishing feedstock. This approach is much more restrictive, but we believe it to be more realistic than assuming perfect sticking up to the fragmentation (or drift) barrier. 
Naturally, slower growth makes overcoming the radial drift barrier even more challenging.}} 

{\postrevisionbf{We also do not adopt the above-cited criteria for the onset of strong acting SI.  While we consider the $\epsilon \gtrsim 1$ criterion to be necessary for strong SI, we generally rely on the more complete solutions and analytical theory of \citetalias{Umurhan_etal_2020}, which successfully back-predicts a multitude of SI simulations in the literature as to whether they lead to SI rapidly growing out of a turbulent state. 
None of our global model cases analyzed in this paper are predicted to become strongly SI active from the analysis of \citetalias{Umurhan_etal_2020}, and (in the sense of leading directly to planetesimals) we never reach those conditions. 
On the other hand, \citetalias{Estrada22_paper2} do find a global model case that develops $\St \gtrsim 0.3$ and $\epsilon > 1$ --
at the snowline and for {\it for small} $\alpha = 10^{-5}$ -- that also happens  {\it to be} consistent with the predictions for strong acting SI according to \citetalias{Umurhan_etal_2020} theory.}} 

{\postrevisionbf{Some other models simulate low gas mass, cold pp-disks \citep[e.g.,][]{Gonzalez_etal_2017,Schoonenberg_etal_2018}, which may be more appropriate for older and significantly evolved Class II objects. Such gas-rarefied environments are generally conducive for strong SI\footnote{\postrevisionbf{It should be noted that in \citet[][see also \citealt{Garcia_Gonzalez2020}]{Gonzalez_etal_2017}, the mechanism proposed to lead to planetesimal formation is self-induced dust traps, and is different from the SI. The dust traps arise due to the combination of growth and fragmentation of dust grains coupled with the back reaction with the gas when $\epsilon = \order{1}$. 
}}. Note that the models of \citet{Carrera2017} begin massive, but they allow photoevaporation to remove disk gas. They find conditions for the SI are met in the outermost and innermost regions of their disk, but only at much later times when gas surface densities have decreased significantly (the local $Z$ and $\St$ increase accordingly) effectively evolving to a Class II pp-disk.}}

{\postrevisionbf{Our model simulations are perhaps most similar to those of \citet[][and also \citealt{HN18,Homma_etal_2019}]{Lenz_etal_2020} who use the full coagulation code of \citet[][also, \citealt{Brauer_etal_2008}]{Birnstiel2010}; however, the main difference is that \citeauthor{Lenz_etal_2020} assume that ``pebble traps'' appear and disappear on some arbitrary timescale within which entrapped pebble clouds can collapse into planetesimals with some specified efficiency\footnote{{\postrevisionbf{This mechanism for planetesimal formation is different from the SI, and might be more akin to turbulent concentration \citep{Hartlep_etal_2020}.}}}. \citet{HN18} modeled the infall stage of the disk with fractal aggregate growth, and found that they could not form planetesimals outside the snow line, whereas \citet{Homma_etal_2019} assumed a sticky organics (``adhesion'') model with a larger fragmentation threshold in the inner disk to produce planetesimals. The peak masses of compact particles and porous aggregates achieved though in these models are similar to those we obtain in our simulations \citep[see also, e.g.,][]{Kri15}.}}

\bibliography{Disk_References}{}

\begin{thebibliography}{}
\expandafter\ifx\csname natexlab\endcsname\relax\def\natexlab#1{#1}\fi
\providecommand{\url}[1]{\href{#1}{#1}}
\providecommand{\dodoi}[1]{doi:~\href{http://doi.org/#1}{\nolinkurl{#1}}}
\providecommand{\doeprint}[1]{\href{http://ascl.net/#1}{\nolinkurl{http://ascl.net/#1}}}
\providecommand{\doarXiv}[1]{\href{https://arxiv.org/abs/#1}{\nolinkurl{https://arxiv.org/abs/#1}}}

\bibitem[{{Alexakis} \& {Biferale}(2018)}]{Alexakis_Biferale_2018}
{Alexakis}, A., \& {Biferale}, L. 2018, \physrep, 767, 1,
  \dodoi{10.1016/j.physrep.2018.08.001}

\bibitem[{{Barranco} {et~al.}(2018){Barranco}, {Pei}, \&
  {Marcus}}]{Barranco_etal_2018}
{Barranco}, J.~A., {Pei}, S., \& {Marcus}, P.~S. 2018, \apj, 869, 127,
  \dodoi{10.3847/1538-4357/aaec80}

\bibitem[{{Birnstiel} {et~al.}(2010){Birnstiel}, {Dullemond}, \&
  {Brauer}}]{Birnstiel2010}
{Birnstiel}, T., {Dullemond}, C.~P., \& {Brauer}, F. 2010, \aap, 513, A79,
  \dodoi{10.1051/0004-6361/200913731}

\bibitem[{{Boffetta} \& {Ecke}(2012)}]{Boffeta_Ecke_2012}
{Boffetta}, G., \& {Ecke}, R.~E. 2012, Annual Review of Fluid Mechanics, 44,
  427, \dodoi{10.1146/annurev-fluid-120710-101240}

\bibitem[{{Brauer} {et~al.}(2008){Brauer}, {Dullemond}, \&
  {Henning}}]{Brauer_etal_2008}
{Brauer}, F., {Dullemond}, C.~P., \& {Henning}, T. 2008, \aap, 480, 859,
  \dodoi{10.1051/0004-6361:20077759}

\bibitem[{{Carrera} {et~al.}(2017){Carrera}, {Gorti}, {Johansen}, \&
  {Davies}}]{Carrera2017}
{Carrera}, D., {Gorti}, U., {Johansen}, A., \& {Davies}, M.~B. 2017, \apj, 839,
  16, \dodoi{10.3847/1538-4357/aa6932}

\bibitem[{{Carrera} {et~al.}(2015){Carrera}, {Johansen}, \&
  {Davies}}]{Carrera_etal_2015}
{Carrera}, D., {Johansen}, A., \& {Davies}, M.~B. 2015, \aap, 579, A43,
  \dodoi{10.1051/0004-6361/201425120}

\bibitem[{{Carrera} \& {Simon}(2022)}]{Carrera_Simon_2022}
{Carrera}, D., \& {Simon}, J.~B. 2022, \apjl, 933, L10,
  \dodoi{10.3847/2041-8213/ac6b3e}

\bibitem[{{Chambers}(2014)}]{Chambers2014}
{Chambers}, J.~E. 2014, \icarus, 233, 83, \dodoi{10.1016/j.icarus.2014.01.036}

\bibitem[{{Charnoz} {et~al.}(2019){Charnoz}, {Pignatale}, {Hyodo}, {Mahan},
  {Chaussidon}, {Siebert}, \& {Moynier}}]{Charnoz_etal_2019}
{Charnoz}, S., {Pignatale}, F.~C., {Hyodo}, R., {et~al.} 2019, \aap, 627, A50,
  \dodoi{10.1051/0004-6361/201833216}

\bibitem[{{Chen} \& {Lin}(2020)}]{Chen_Lin_2020}
{Chen}, K., \& {Lin}, M.-K. 2020, \apj, 891, 132,
  \dodoi{10.3847/1538-4357/ab76ca}

\bibitem[{{Connelly} {et~al.}(2012){Connelly}, {Bizzarro}, {Krot}, {Nordlund},
  {Wielandt}, \& {Ivanova}}]{Connelly2012}
{Connelly}, J.~N., {Bizzarro}, M., {Krot}, A.~N., {et~al.} 2012, Science, 338,
  651, \dodoi{10.1126/science.1226919}

\bibitem[{{Cuzzi} {et~al.}(1993){Cuzzi}, {Dobrovolskis}, \&
  {Champney}}]{Cuzzi1993}
{Cuzzi}, J.~N., {Dobrovolskis}, A.~R., \& {Champney}, J.~M. 1993, \icarus, 106,
  102, \dodoi{10.1006/icar.1993.1161}

\bibitem[{{D'Angelo} {et~al.}(2003){D'Angelo}, {Henning}, \&
  {Kley}}]{Dangelo_etal_2003}
{D'Angelo}, G., {Henning}, T., \& {Kley}, W. 2003, \apj, 599, 548,
  \dodoi{10.1086/379224}

\bibitem[{{Desch}(2007)}]{Des07}
{Desch}, S.~J. 2007, \apj, 671, 878, \dodoi{10.1086/522825}

\bibitem[{{Desch} {et~al.}(2018){Desch}, {Kalyaan}, \& {O'D.
  Alexander}}]{Desch_etal_2018}
{Desch}, S.~J., {Kalyaan}, A., \& {O'D. Alexander}, C.~M. 2018, \apjs, 238, 11,
  \dodoi{10.3847/1538-4365/aad95f}

\bibitem[{{Dra{\.z}kowska} \& {Alibert}(2017)}]{Draz_Alibert2017}
{Dra{\.z}kowska}, J., \& {Alibert}, Y. 2017, \aap, 608, A92,
  \dodoi{10.1051/0004-6361/201731491}

\bibitem[{{Dra{\.z}kowska} {et~al.}(2016){Dra{\.z}kowska}, {Alibert}, \&
  {Moore}}]{Draz2016}
{Dra{\.z}kowska}, J., {Alibert}, Y., \& {Moore}, B. 2016, \aap, 594, A105,
  \dodoi{10.1051/0004-6361/201628983}

\bibitem[{{Dra{\.z}kowska} \& {Dullemond}(2018)}]{DD18}
{Dra{\.z}kowska}, J., \& {Dullemond}, C.~P. 2018, \aap, 614, A62,
  \dodoi{10.1051/0004-6361/201732221}

\bibitem[{{Dra{\.z}kowska} {et~al.}(2019){Dra{\.z}kowska}, {Li}, {Birnstiel},
  {Stammler}, \& {Li}}]{Draz2019}
{Dra{\.z}kowska}, J., {Li}, S., {Birnstiel}, T., {Stammler}, S.~M., \& {Li}, H.
  2019, \apj, 885, 91, \dodoi{10.3847/1538-4357/ab46b7}

\bibitem[{{Dubrulle} {et~al.}(1995){Dubrulle}, {Morfill}, \&
  {Sterzik}}]{Dubrulle_etal_1995}
{Dubrulle}, B., {Morfill}, G., \& {Sterzik}, M. 1995, \icarus, 114, 237,
  \dodoi{10.1006/icar.1995.1058}

\bibitem[{{Dullemond} {et~al.}(2018){Dullemond}, {Birnstiel}, {Huang},
  {Kurtovic}, {Andrews}, {Guzm{\'a}n}, {P{\'e}rez}, {Isella}, {Zhu}, {Benisty},
  {Wilner}, {Bai}, {Carpenter}, {Zhang}, \& {Ricci}}]{Dullemond2018}
{Dullemond}, C.~P., {Birnstiel}, T., {Huang}, J., {et~al.} 2018, \apjl, 869,
  L46, \dodoi{10.3847/2041-8213/aaf742}

\bibitem[{{Estrada} \& {Cuzzi}(2022)}]{Estrada22_paper3}
{Estrada}, P.~R., \& {Cuzzi}, J.~N. 2022, \apj, 936, 40,
  \dodoi{10.3847/1538-4357/ac81c6}

\bibitem[{{Estrada} {et~al.}(2016){Estrada}, {Cuzzi}, \&
  {Morgan}}]{Estrada_etal_2016}
{Estrada}, P.~R., {Cuzzi}, J.~N., \& {Morgan}, D.~A. 2016, \apj, 818, 200,
  \dodoi{10.3847/0004-637X/818/2/200}

\bibitem[{{Estrada} {et~al.}(2022){Estrada}, {Cuzzi}, \&
  {Umurhan}}]{Estrada22_paper2}
{Estrada}, P.~R., {Cuzzi}, J.~N., \& {Umurhan}, O.~M. 2022, \apj, 936, 42,
  \dodoi{10.3847/1538-4357/ac7ffd}

\bibitem[{{Flaherty} {et~al.}(2018){Flaherty}, {Hughes}, {Teague}, {Simon},
  {Andrews}, \& {Wilner}}]{Flaherty2018}
{Flaherty}, K.~M., {Hughes}, A.~M., {Teague}, R., {et~al.} 2018, \apj, 856,
  117, \dodoi{10.3847/1538-4357/aab615}

\bibitem[{{Flaherty} {et~al.}(2017){Flaherty}, {Hughes}, {Rose}, {Simon}, {Qi},
  {Andrews}, {K{\'o}sp{\'a}l}, {Wilner}, {Chiang}, {Armitage}, \&
  {Bai}}]{Flaherty2017}
{Flaherty}, K.~M., {Hughes}, A.~M., {Rose}, S.~C., {et~al.} 2017, \apj, 843,
  150, \dodoi{10.3847/1538-4357/aa79f9}

\bibitem[{{Flock} {et~al.}(2017){Flock}, {Nelson}, {Turner}, {Bertrang},
  {Carrasco-Gonz{\'a}lez}, {Henning}, {Lyra}, \& {Teague}}]{Flock_etal_2017}
{Flock}, M., {Nelson}, R.~P., {Turner}, N.~J., {et~al.} 2017, \apj, 850, 131,
  \dodoi{10.3847/1538-4357/aa943f}

\bibitem[{{Flock} {et~al.}(2020){Flock}, {Turner}, {Nelson}, {Lyra}, {Manger},
  \& {Klahr}}]{Flock_etal_2020}
{Flock}, M., {Turner}, N.~J., {Nelson}, R.~P., {et~al.} 2020, \apj, 897, 155,
  \dodoi{10.3847/1538-4357/ab9641}

\bibitem[{{Garcia} \& {Gonzalez}(2020)}]{Garcia_Gonzalez2020}
{Garcia}, A. J.~L., \& {Gonzalez}, J.-F. 2020, \mnras, 493, 1788,
  \dodoi{10.1093/mnras/staa382}

\bibitem[{{Gole} {et~al.}(2020){Gole}, {Simon}, {Li}, {Youdin}, \&
  {Armitage}}]{Gole_etal_2020}
{Gole}, D.~A., {Simon}, J.~B., {Li}, R., {Youdin}, A.~N., \& {Armitage}, P.~J.
  2020, \apj, 904, 132, \dodoi{10.3847/1538-4357/abc334}

\bibitem[{{Gonzalez} {et~al.}(2017){Gonzalez}, {Laibe}, \&
  {Maddison}}]{Gonzalez_etal_2017}
{Gonzalez}, J.~F., {Laibe}, G., \& {Maddison}, S.~T. 2017, \mnras, 467, 1984,
  \dodoi{10.1093/mnras/stx016}

\bibitem[{{G{\"u}ttler} {et~al.}(2010){G{\"u}ttler}, {Blum}, {Zsom}, {Ormel},
  \& {Dullemond}}]{Guttler_etal_2010}
{G{\"u}ttler}, C., {Blum}, J., {Zsom}, A., {Ormel}, C.~W., \& {Dullemond},
  C.~P. 2010, \aap, 513, A56, \dodoi{10.1051/0004-6361/200912852}

\bibitem[{{Hartlep} \& {Cuzzi}(2020)}]{Hartlep_etal_2020}
{Hartlep}, T., \& {Cuzzi}, J.~N. 2020, \apj, 892, 120,
  \dodoi{10.3847/1538-4357/ab76c3}

\bibitem[{{Hartmann} {et~al.}(1998){Hartmann}, {Calvet}, {Gullbring}, \&
  {D'Alessio}}]{Har98}
{Hartmann}, L., {Calvet}, N., {Gullbring}, E., \& {D'Alessio}, P. 1998, \apj,
  495, 385, \dodoi{10.1086/305277}

\bibitem[{{Homma} \& {Nakamoto}(2018)}]{HN18}
{Homma}, K., \& {Nakamoto}, T. 2018, \apj, 868, 118,
  \dodoi{10.3847/1538-4357/aae0fb}

\bibitem[{{Homma} {et~al.}(2019){Homma}, {Okuzumi}, {Nakamoto}, \&
  {Ueda}}]{Homma_etal_2019}
{Homma}, K.~A., {Okuzumi}, S., {Nakamoto}, T., \& {Ueda}, Y. 2019, \apj, 877,
  128, \dodoi{10.3847/1538-4357/ab1de0}

\bibitem[{{Jacquet} {et~al.}(2019){Jacquet}, {Pignatale}, {Chaussidon}, \&
  {Charnoz}}]{Jacquet2019}
{Jacquet}, E., {Pignatale}, F.~C., {Chaussidon}, M., \& {Charnoz}, S. 2019,
  \apj, 884, 32, \dodoi{10.3847/1538-4357/ab38c1}

\bibitem[{{Joswiak} {et~al.}(2012){Joswiak}, {Brownlee}, {Matrajt}, {Westphal},
  {Snead}, \& {Gainsforth}}]{Joswiak2012}
{Joswiak}, D.~J., {Brownlee}, D.~E., {Matrajt}, G., {et~al.} 2012, \maps, 47,
  471, \dodoi{10.1111/j.1945-5100.2012.01337.x}

\bibitem[{{Kataoka} {et~al.}(2013){Kataoka}, {Tanaka}, {Okuzumi}, \&
  {Wada}}]{Kat13a}
{Kataoka}, A., {Tanaka}, H., {Okuzumi}, S., \& {Wada}, K. 2013, \aap, 554, A4,
  \dodoi{10.1051/0004-6361/201321325}

\bibitem[{{Kita} \& {Ushikubo}(2012)}]{Kita2012}
{Kita}, N.~T., \& {Ushikubo}, T. 2012, \maps, 47, 1108,
  \dodoi{10.1111/j.1945-5100.2011.01264.x}

\bibitem[{{Kita} {et~al.}(2013){Kita}, {Yin}, {MacPherson}, {Ushikubo},
  {Jacobsen}, {Nagashima}, {Kurahashi}, {Krot}, \& {Jacobsen}}]{Kita2013}
{Kita}, N.~T., {Yin}, Q.-Z., {MacPherson}, G.~J., {et~al.} 2013, \maps, 48,
  1383, \dodoi{10.1111/maps.12141}

\bibitem[{{Krijt} {et~al.}(2015){Krijt}, {Ormel}, {Dominik}, \&
  {Tielens}}]{Kri15}
{Krijt}, S., {Ormel}, C.~W., {Dominik}, C., \& {Tielens}, A.~G.~G.~M. 2015,
  \aap, 574, A83, \dodoi{10.1051/0004-6361/201425222}

\bibitem[{{Krot} {et~al.}(2009){Krot}, {Amelin}, {Bland}, {Ciesla}, {Connelly},
  {Davis}, {Huss}, {Hutcheon}, {Makide}, {Nagashima}, {Nyquist}, {Russell},
  {Scott}, {Thrane}, {Yurimoto}, \& {Yin}}]{Krot2009}
{Krot}, A.~N., {Amelin}, Y., {Bland}, P., {et~al.} 2009, \gca, 73, 4963,
  \dodoi{10.1016/j.gca.2008.09.039}

\bibitem[{{Kruijer} {et~al.}(2017){Kruijer}, {Burkhardt}, {Budde}, \&
  {Kleine}}]{Kruijer2017}
{Kruijer}, T.~S., {Burkhardt}, C., {Budde}, G., \& {Kleine}, T. 2017,
  Proceedings of the National Academy of Science, 114, 6712,
  \dodoi{10.1073/pnas.1704461114}

\bibitem[{{Lambrechts} \& {Johansen}(2012)}]{Lambrechts2012}
{Lambrechts}, M., \& {Johansen}, A. 2012, \aap, 544, A32,
  \dodoi{10.1051/0004-6361/201219127}

\bibitem[{{Lehmann} \& {Lin}(2022)}]{Lehmann_Lin_2022}
{Lehmann}, M., \& {Lin}, M.~K. 2022, \aap, 658, A156,
  \dodoi{10.1051/0004-6361/202142378}

\bibitem[{{Lenz} {et~al.}(2020){Lenz}, {Klahr}, {Birnstiel}, {Kretke}, \&
  {Stammler}}]{Lenz_etal_2020}
{Lenz}, C.~T., {Klahr}, H., {Birnstiel}, T., {Kretke}, K., \& {Stammler}, S.
  2020, \aap, 640, A61, \dodoi{10.1051/0004-6361/202037878}

\bibitem[{{Lesur} {et~al.}(2022){Lesur}, {Ercolano}, {Flock}, {Lin}, {Yang},
  {Barranco}, {Benitez-Llambay}, {Goodman}, {Johansen}, {Klahr}, {Laibe},
  {Lyra}, {Marcus}, {Nelson}, {Squire}, {Simon}, {Turner}, {Umurhan}, \&
  {Youdin}}]{Lesur_etal_2022}
{Lesur}, G., {Ercolano}, B., {Flock}, M., {et~al.} 2022, arXiv e-prints,
  arXiv:2203.09821.
\newblock \doarXiv{2203.09821}

\bibitem[{{Li} \& {Youdin}(2021)}]{Li_Youdin_2021}
{Li}, R., \& {Youdin}, A.~N. 2021, \apj, 919, 107,
  \dodoi{10.3847/1538-4357/ac0e9f}

\bibitem[{{Lin} \& {Papaloizou}(1979)}]{LP79}
{Lin}, D.~N.~C., \& {Papaloizou}, J. 1979, \mnras, 186, 799,
  \dodoi{10.1093/mnras/186.4.799}

\bibitem[{{Lin}(2019)}]{Lin_2019}
{Lin}, M.-K. 2019, \mnras, 485, 5221, \dodoi{10.1093/mnras/stz701}

\bibitem[{{Lynden-Bell} \& {Pringle}(1974)}]{LP74}
{Lynden-Bell}, D., \& {Pringle}, J.~E. 1974, \mnras, 168, 603,
  \dodoi{10.1093/mnras/168.3.603}

\bibitem[{{Lyra} \& {Umurhan}(2019)}]{Lyra_Umurhan_2019}
{Lyra}, W., \& {Umurhan}, O.~M. 2019, \pasp, 131, 072001,
  \dodoi{10.1088/1538-3873/aaf5ff}

\bibitem[{{Manger} {et~al.}(2021){Manger}, {Pfeil}, \&
  {Klahr}}]{Manger_etal_2021}
{Manger}, N., {Pfeil}, T., \& {Klahr}, H. 2021, \mnras, 508, 5402,
  \dodoi{10.1093/mnras/stab2599}

\bibitem[{{Marcus} {et~al.}(2016){Marcus}, {Pei}, {Jiang}, \&
  {Barranco}}]{Marcus_etal_2016}
{Marcus}, P.~S., {Pei}, S., {Jiang}, C.-H., \& {Barranco}, J.~A. 2016, \apj,
  833, 148, \dodoi{10.3847/1538-4357/833/2/148}

\bibitem[{{Marcus} {et~al.}(2013){Marcus}, {Pei}, {Jiang}, \&
  {Hassanzadeh}}]{Marcus_etal_2013}
{Marcus}, P.~S., {Pei}, S., {Jiang}, C.-H., \& {Hassanzadeh}, P. 2013, \prl,
  111, 084501, \dodoi{10.1103/PhysRevLett.111.084501}

\bibitem[{{Marrocchi} {et~al.}(2019){Marrocchi}, {Villeneuve}, {Jacquet},
  {Piralla}, \& {Chaussidon}}]{Marrocchi2019}
{Marrocchi}, Y., {Villeneuve}, J., {Jacquet}, E., {Piralla}, M., \&
  {Chaussidon}, M. 2019, Proceedings of the National Academy of Science, 116,
  23461, \dodoi{10.1073/pnas.1912479116}

\bibitem[{{Morbidelli} {et~al.}(2022){Morbidelli}, {Bailli{\'e}}, {Batygin},
  {Charnoz}, {Guillot}, {Rubie}, \& {Kleine}}]{Morbidelli_etal_2022}
{Morbidelli}, A., {Bailli{\'e}}, K., {Batygin}, K., {et~al.} 2022, Nature
  Astronomy, 6, 72, \dodoi{10.1038/s41550-021-01517-7}

\bibitem[{{Morbidelli} {et~al.}(2009){Morbidelli}, {Bottke}, {Nesvorn{\'y}}, \&
  {Levison}}]{Morbidelli2009}
{Morbidelli}, A., {Bottke}, W.~F., {Nesvorn{\'y}}, D., \& {Levison}, H.~F.
  2009, \icarus, 204, 558, \dodoi{10.1016/j.icarus.2009.07.011}

\bibitem[{{Nakagawa} {et~al.}(1986){Nakagawa}, {Sekiya}, \&
  {Hayashi}}]{Nakagawa_etal_1986}
{Nakagawa}, Y., {Sekiya}, M., \& {Hayashi}, C. 1986, \icarus, 67, 375,
  \dodoi{10.1016/0019-1035(86)90121-1}

\bibitem[{{Nanne} {et~al.}(2019){Nanne}, {Nimmo}, {Cuzzi}, \&
  {Kleine}}]{Nanne2019}
{Nanne}, J. A.~M., {Nimmo}, F., {Cuzzi}, J.~N., \& {Kleine}, T. 2019, Earth and
  Planetary Science Letters, 511, 44, \dodoi{10.1016/j.epsl.2019.01.027}

\bibitem[{{National Academies of Sciences, Engineering, and
  Medicine}(2022)}]{Planetary_Science_Decadal_Survey_2022}
{National Academies of Sciences, Engineering, and Medicine}. 2022, {Origins,
  Worlds, Life: A Decadal Strategy for Planetary Science and Astrobiology
  2023-2032.}

\bibitem[{{Nesvorn{\'y}} {et~al.}(2019){Nesvorn{\'y}}, {Li}, {Youdin}, {Simon},
  \& {Grundy}}]{Nesvorny_etal_2019}
{Nesvorn{\'y}}, D., {Li}, R., {Youdin}, A.~N., {Simon}, J.~B., \& {Grundy},
  W.~M. 2019, Nature Astronomy, 3, 808, \dodoi{10.1038/s41550-019-0806-z}

\bibitem[{{Okuzumi} \& {Hirose}(2012)}]{Okuzumi_Hirose2012}
{Okuzumi}, S., \& {Hirose}, S. 2012, \apjl, 753, L8,
  \dodoi{10.1088/2041-8205/753/1/L8}

\bibitem[{{Okuzumi} {et~al.}(2012){Okuzumi}, {Tanaka}, {Kobayashi}, \&
  {Wada}}]{Oku12}
{Okuzumi}, S., {Tanaka}, H., {Kobayashi}, H., \& {Wada}, K. 2012, \apj, 752,
  106, \dodoi{10.1088/0004-637X/752/2/106}

\bibitem[{{Ormel} \& {Klahr}(2010)}]{Ormel2010}
{Ormel}, C.~W., \& {Klahr}, H.~H. 2010, \aap, 520, A43,
  \dodoi{10.1051/0004-6361/201014903}

\bibitem[{{Paardekooper} \& {Johansen}(2018)}]{Paardekooper2018}
{Paardekooper}, S.-J., \& {Johansen}, A. 2018, \ssr, 214, 38,
  \dodoi{10.1007/s11214-018-0472-y}

\bibitem[{{Paardekooper} \& {Mellema}(2006)}]{Paardekooper2006}
{Paardekooper}, S.~J., \& {Mellema}, G. 2006, \aap, 453, 1129,
  \dodoi{10.1051/0004-6361:20054449}

\bibitem[{{Raettig} {et~al.}(2021){Raettig}, {Lyra}, \&
  {Klahr}}]{Raettig_etal_2021}
{Raettig}, N., {Lyra}, W., \& {Klahr}, H. 2021, \apj, 913, 92,
  \dodoi{10.3847/1538-4357/abf739}

\bibitem[{{Rice} {et~al.}(2006){Rice}, {Armitage}, {Wood}, \&
  {Lodato}}]{Rice2006}
{Rice}, W.~K.~M., {Armitage}, P.~J., {Wood}, K., \& {Lodato}, G. 2006, \mnras,
  373, 1619, \dodoi{10.1111/j.1365-2966.2006.11113.x}

\bibitem[{{Richard} {et~al.}(2016){Richard}, {Nelson}, \&
  {Umurhan}}]{Richard_etal_2016}
{Richard}, S., {Nelson}, R.~P., \& {Umurhan}, O.~M. 2016, \mnras, 456, 3571,
  \dodoi{10.1093/mnras/stv2898}

\bibitem[{{Sch{\"a}fer} \& {Johansen}(2022)}]{Schafer_Johansen_2022}
{Sch{\"a}fer}, U., \& {Johansen}, A. 2022, \aap, 666, A98,
  \dodoi{10.1051/0004-6361/202243655}

\bibitem[{{Sch{\"a}fer} {et~al.}(2020){Sch{\"a}fer}, {Johansen}, \&
  {Banerjee}}]{Schafer_etal_2020}
{Sch{\"a}fer}, U., {Johansen}, A., \& {Banerjee}, R. 2020, \aap, 635, A190,
  \dodoi{10.1051/0004-6361/201937371}

\bibitem[{{Schoonenberg} {et~al.}(2018){Schoonenberg}, {Ormel}, \&
  {Krijt}}]{Schoonenberg_etal_2018}
{Schoonenberg}, D., {Ormel}, C.~W., \& {Krijt}, S. 2018, \aap, 620, A134,
  \dodoi{10.1051/0004-6361/201834047}

\bibitem[{{Schrader} {et~al.}(2017){Schrader}, {Nagashima}, {Krot}, {Ogliore},
  {Yin}, {Amelin}, {Stirling}, \& {Kaltenbach}}]{Schrader2017}
{Schrader}, D.~L., {Nagashima}, K., {Krot}, A.~N., {et~al.} 2017, \gca, 201,
  275, \dodoi{10.1016/j.gca.2016.06.023}

\bibitem[{{Sekiya} \& {Onishi}(2018)}]{Sekiya_Onishi_2018}
{Sekiya}, M., \& {Onishi}, I.~K. 2018, \apj, 860, 140,
  \dodoi{10.3847/1538-4357/aac4a7}

\bibitem[{{Sengupta} {et~al.}(2022){Sengupta}, {Estrada}, {Cuzzi}, \&
  {Humayun}}]{Sengupta2022}
{Sengupta}, D., {Estrada}, P.~R., {Cuzzi}, J.~N., \& {Humayun}, M. 2022, \apj,
  932, 82, \dodoi{10.3847/1538-4357/ac6dcc}

\bibitem[{{Sengupta} \& {Umurhan}(2023)}]{Sengupta_Umurhan2023}
{Sengupta}, D., \& {Umurhan}, O.~M. 2023, \apj, 942, 74,
  \dodoi{10.3847/1538-4357/ac9411}

\bibitem[{{Simon} {et~al.}(2017){Simon}, {Armitage}, {Youdin}, \&
  {Li}}]{Simon_etal_2017}
{Simon}, J.~B., {Armitage}, P.~J., {Youdin}, A.~N., \& {Li}, R. 2017, \apjl,
  847, L12, \dodoi{10.3847/2041-8213/aa8c79}

\bibitem[{{Simon} {et~al.}(2019){Simon}, {Ross}, {Nguyen}, {Simon}, \&
  {Messenger}}]{Simon2019}
{Simon}, J.~I., {Ross}, D.~K., {Nguyen}, A.~N., {Simon}, S.~B., \& {Messenger},
  S. 2019, \apjl, 884, L29, \dodoi{10.3847/2041-8213/ab43e4}

\bibitem[{{Squire} \& {Hopkins}(2018{\natexlab{a}})}]{Squire_Hopkins_2018a}
{Squire}, J., \& {Hopkins}, P.~F. 2018{\natexlab{a}}, \mnras, 477, 5011,
  \dodoi{10.1093/mnras/sty854}

\bibitem[{{Squire} \& {Hopkins}(2018{\natexlab{b}})}]{Squire_Hopkins_2018b}
---. 2018{\natexlab{b}}, \apjl, 856, L15, \dodoi{10.3847/2041-8213/aab54d}

\bibitem[{{Stoll} {et~al.}(2017){Stoll}, {Kley}, \&
  {Picogna}}]{Stoll_Kley_Picogna_2017}
{Stoll}, M. H.~R., {Kley}, W., \& {Picogna}, G. 2017, \aap, 599, L6,
  \dodoi{10.1051/0004-6361/201630226}

\bibitem[{{Suyama} {et~al.}(2012){Suyama}, {Wada}, {Tanaka}, \&
  {Okuzumi}}]{Suy12}
{Suyama}, T., {Wada}, K., {Tanaka}, H., \& {Okuzumi}, S. 2012, \apj, 753, 115,
  \dodoi{10.1088/0004-637X/753/2/115}

\bibitem[{{Teague} {et~al.}(2016){Teague}, {Guilloteau}, {Semenov}, {Henning},
  {Dutrey}, {Pi{\'e}tu}, {Birnstiel}, {Chapillon}, {Hollenbach}, \&
  {Gorti}}]{Teague2016}
{Teague}, R., {Guilloteau}, S., {Semenov}, D., {et~al.} 2016, \aap, 592, A49,
  \dodoi{10.1051/0004-6361/201628550}

\bibitem[{{Turner} {et~al.}(2014){Turner}, {Fromang}, {Gammie}, {Klahr},
  {Lesur}, {Wardle}, \& {Bai}}]{Turner2014}
{Turner}, N.~J., {Fromang}, S., {Gammie}, C., {et~al.} 2014, in Protostars and
  Planets VI, ed. H.~{Beuther}, R.~S. {Klessen}, C.~P. {Dullemond}, \&
  T.~{Henning}, 411, \dodoi{10.2458/azu_uapress_9780816531240-ch018}

\bibitem[{{Umurhan} {et~al.}(2020){Umurhan}, {Estrada}, \&
  {Cuzzi}}]{Umurhan_etal_2020}
{Umurhan}, O.~M., {Estrada}, P.~R., \& {Cuzzi}, J.~N. 2020, \apj, 895, 4,
  \dodoi{10.3847/1538-4357/ab899d}

\bibitem[{{Umurhan} {et~al.}(2016){Umurhan}, {Nelson}, \&
  {Gressel}}]{Umurhan_etal_2016}
{Umurhan}, O.~M., {Nelson}, R.~P., \& {Gressel}, O. 2016, \aap, 586, A33,
  \dodoi{10.1051/0004-6361/201526494}

\bibitem[{{Urpin}(2003)}]{urpin03}
{Urpin}, V. 2003, \aap, 404, 397, \dodoi{10.1051/0004-6361:20030513}

\bibitem[{{Yang} {et~al.}(2017){Yang}, {Johansen}, \&
  {Carrera}}]{Yang_etal_2017}
{Yang}, C.-C., {Johansen}, A., \& {Carrera}, D. 2017, \aap, 606, A80,
  \dodoi{10.1051/0004-6361/201630106}

\bibitem[{{Yellin-Bergovoy} {et~al.}(2021){Yellin-Bergovoy}, {Heifetz}, \&
  {Umurhan}}]{Yellin-Bergovoy_etal_2021}
{Yellin-Bergovoy}, R., {Heifetz}, E., \& {Umurhan}, O.~M. 2021, arXiv e-prints,
  arXiv:2106.04617.
\newblock \doarXiv{2106.04617}

\bibitem[{{Youdin} \& {Goodman}(2005)}]{Youdin_Goodman_2005}
{Youdin}, A.~N., \& {Goodman}, J. 2005, \apj, 620, 459, \dodoi{10.1086/426895}

\bibitem[{{Zanda} {et~al.}(2018){Zanda}, {Lewin}, \& {Humayun}}]{Zanda2018}
{Zanda}, B., {Lewin}, E., \& {Humayun}, M. 2018, in Chondrules: Records of
  Protoplanetary Disk Processes, ed. S.~S. {Russell}, J.~{Connolly}, Harold~C.,
  \& A.~N. {Krot}, 122--150, \dodoi{10.1017/9781108284073.005}

\bibitem[{{Zolensky} {et~al.}(2006){Zolensky}, {Zega}, {Yano}, {Wirick},
  {Westphal}, {Weisberg}, {Weber}, {Warren}, {Velbel}, {Tsuchiyama}, {Tsou},
  {Toppani}, {Tomioka}, {Tomeoka}, {Teslich}, {Taheri}, {Susini}, {Stroud},
  {Stephan}, {Stadermann}, {Snead}, {Simon}, {Simionovici}, {See}, {Robert},
  {Rietmeijer}, {Rao}, {Perronnet}, {Papanastassiou}, {Okudaira}, {Ohsumi},
  {Ohnishi}, {Nakamura-Messenger}, {Nakamura}, {Mostefaoui}, {Mikouchi},
  {Meibom}, {Matrajt}, {Marcus}, {Leroux}, {Lemelle}, {Le}, {Lanzirotti},
  {Langenhorst}, {Krot}, {Keller}, {Kearsley}, {Joswiak}, {Jacob}, {Ishii},
  {Harvey}, {Hagiya}, {Grossman}, {Grossman}, {Graham}, {Gounelle}, {Gillet},
  {Genge}, {Flynn}, {Ferroir}, {Fallon}, {Ebel}, {Dai}, {Cordier}, {Clark},
  {Chi}, {Butterworth}, {Brownlee}, {Bridges}, {Brennan}, {Brearley},
  {Bradley}, {Bleuet}, {Bland}, \& {Bastien}}]{Zolensky2006}
{Zolensky}, M.~E., {Zega}, T.~J., {Yano}, H., {et~al.} 2006, Science, 314,
  1735, \dodoi{10.1126/science.1135842}

\bibitem[{{Zsom} {et~al.}(2010){Zsom}, {Ormel}, {G{\"u}ttler}, {Blum}, \&
  {Dullemond}}]{Zsom_etal_2010}
{Zsom}, A., {Ormel}, C.~W., {G{\"u}ttler}, C., {Blum}, J., \& {Dullemond},
  C.~P. 2010, \aap, 513, A57, \dodoi{10.1051/0004-6361/200912976}

\end{thebibliography}
\bibliographystyle{aasjournal}

\end{document}